\documentclass[fleqn,usenatbib]{mnras}
\usepackage{newtxtext,newtxmath}
\usepackage[T1]{fontenc}

\DeclareRobustCommand{\VAN}[3]{#2}
\let\VANthebibliography\thebibliography
\def\thebibliography{\DeclareRobustCommand{\VAN}[3]{##3}\VANthebibliography}

\usepackage{graphicx}
\usepackage{dcolumn}
\usepackage{bm}
\usepackage{tabularx}
\usepackage{booktabs}
\usepackage{hyperref}
\usepackage{physics}
\usepackage{cancel}
\usepackage{enumitem}
\usepackage{amsmath}
\usepackage{iftex}
\ifxetex
  \usepackage{fontspec}
  \usepackage{xeCJK}
\fi
\usepackage{xcolor}

\newcommand{\LyA}{Lyman-$\alpha$~}
\newcommand{\fA}{\ensuremath f_\text{A}}
\newcommand{\axionyx}{\texttt{AxioNyx}}
\newcommand{\maxi}{\ensuremath m_\text{A}~}

\ifxetex
  \newcommand{\AuthorNameLine}{Yourong~Frank~Wang~(王有容)}
\else
  \newcommand{\AuthorNameLine}{Yourong~Frank~Wang}
\fi

\title[Ly-$\alpha$ Forest of Mixed Fuzzy and Cold Dark Matter]{\LyA Forest Signatures of Mixed Fuzzy and Cold Dark Matter}

\author[Y. F. Wang]{
\AuthorNameLine\thanks{E-mail: yourong.wang@uni-goettingen.de} \\
Institut f\"ur Astrophysik, Georg-August-Universit\"at 
 Göttingen, D-37077 Göttingen, Germany\\
}

\date{Accepted XXX. Received YYY; in original form ZZZ}

\pubyear{\the\year{}}

\definecolor{revcolor}{rgb}{0,0,0}

\newcommand{\revstart}{\begingroup\color{revcolor}}
\newcommand{\revend}{\endgroup}
\begin{document}
\label{firstpage}
\pagerange{\pageref{firstpage}--\pageref{lastpage}}
\maketitle

\begin{abstract}
\revstart We investigate \LyA forest flux statistics in mixed fuzzy dark matter (FDM) and cold dark matter (CDM) cosmologies using the Fluctuating Gunn--Peterson Approximation (FGPA) applied to hybrid Schr\"odinger--Poisson and $N$-body simulations. We evolve the dark matter distribution from $z = 120$ to $z = 2$ for an axion mass $\maxi = 10^{-24}\,\text{eV}$ and FDM fraction $f_{\rm A} = 0.1$, and compare three simulations sharing random phases: a $\Lambda$CDM baseline, an MDM realisation evolved with full wave-mechanical (Schr\"odinger--Poisson) dynamics, and a particle-only ($N$-body) MDM proxy with matched initial total-matter density power and CDM-like initial velocities. Because the initial velocity prescriptions differ, the comparison combines an inherited velocity offset with the subsequent dynamical evolution. By $z=4$ the nonlinear matter power spectra of the two MDM treatments differ by only $\sim25$ per cent at $k\sim10\,h\,\mathrm{Mpc}^{-1}$, whereas their (unfiltered) line-of-sight velocity power spectra differ more at the same scale, the full treatment retaining $\sim44$ per cent of the $N$-body velocity power. The corresponding \LyA flux power spectra differ by $21$ per cent at $z = 4$ and $3$ per cent at $z = 2$ at $k = 0.1\,\mathrm{s\,km^{-1}}$, with the difference growing toward higher redshift and smaller scales. The flux statistics therefore depend on the velocity prescription and not only on the matter power spectrum, and a particle-only proxy based on suppressed initial density power does not reproduce them within this framework. The analysis is idealised: it is dark-matter-only, relies on the FGPA with a prescribed thermal state, and uses only a preliminary neutral-hydrogen-weighted velocity diagnostic. Establishing whether the difference survives baryonic response and thermal-history uncertainty requires full hydrodynamical simulations. \revend
\end{abstract}


\section{Introduction}

The \LyA forest, absorption features due to neutral hydrogen in the intergalactic medium (IGM), provides one of the most sensitive probes of small-scale structure at redshifts $2 < z < 6$ \citep{Hernquist96}, constraining the clustering of matter on comoving scales of order $1\text{--}10\,h\,\mathrm{Mpc}^{-1}$ \citep{Irsic2017}.

\revstart Models of fuzzy dark matter (FDM / ULDM) have been discussed as a possible alternative to cold dark matter (CDM), potentially modifying structure formation on kiloparsec scales through wave-mechanical effects \citep{Hu2000, Hui2017, NiemeyerR2020, May2022, Eberhardt2025}. The \LyA signatures of FDM have been studied through observational analyses based largely on hydrodynamical or \(N\)-body simulations with FDM-suppressed initial conditions \citep{Armengaud2017,Irsic2017,RogersPeiris2021}, through particle-based effective quantum-pressure treatments \citep{Nori2018,NoriMNRAS}, and in direct Schrödinger--Poisson calculations \citep{Li2019}. Full wave-mechanical cosmological simulations have also been used to study nonlinear FDM structure formation more generally \citep{Veltmaat2018,May2022}, and the Schrödinger--Poisson formalism has also been used to describe the evolution of cosmic voids \citep{gallagher_coles_2022}. \revend

\revstart A particle-based treatment incorporating an effective quantum-pressure force was developed by \citet{Nori2018} as the \textsc{Ax-Gadget} code, in which the FDM quantum pressure enters as an additional force computed from the density estimate, alongside a full hydrodynamic treatment of the baryons. The Madelung correspondence underlying such treatments becomes singular at nodes of the wavefunction and does not evolve the phase information sufficiently robustly to follow the full interference dynamics in the nonlinear regime. Further, \citet{Li2019} directly compared Schrödinger--Poisson and $N$-body evolution from matched pure-FDM initial conditions, finding that, at a redshift of $z=5$ and for $\maxi \gtrsim 2\times10^{-23}$eV, the \LyA flux power spectra agree to better than 10 per cent at scales $k \lesssim 0.1 \text{s km}^{-1}$ between the methods. Their calculation provided an early direct test of the extent to which suppressed initial conditions evolved with $N$-body dynamics can approximate a fully resolved wave simulation. \revend

\revstart Physically motivated cosmologies may, however, consist of a mixture of FDM and cold dark matter (CDM), arising naturally in multi-field ultralight dark matter models when one species is sufficiently massive to behave effectively as CDM \citep{Multifield}. How the comparison between wave and $N$-body evolution changes in this regime is less clear: the FDM component may exhibit strong wave dynamics while the CDM-dominated total density field remains comparatively insensitive. In an analytic linear-theory treatment, \citet{Kobayashi2017} found that the suppression of the total matter power is strongly diluted for subdominant FDM: for $f_{\rm A}\lesssim0.07$, their estimate gives no factor-of-two suppression of the present-day linear power on any scale, leading them to conclude that FDM has little impact on structure formation for $f_{\rm A}\lesssim0.1$, essentially independently of the axion mass. Their full hydrodynamical \LyA analysis instead found that the forest becomes practically insensitive for $f_{\rm A}\lesssim0.2$. Their analysis, like the subsequent mixed-model constraint of \citet{Liu2026MDM}, does not include quantum pressure during the nonlinear evolution; \citet{Kobayashi2017} argued that the resulting bounds should therefore be regarded as conservative, since nonlinear quantum pressure could introduce additional suppression. On the perturbative side, the effective field theory of large-scale structure has recently been extended to mixed dark matter, describing the non-cold component as a fluid with finite sound speed \citep{Verd2025}; such a description does not follow wave interference in the nonlinear regime, which is the regime probed here. \revend

We address this gap using \axionyx\ \citep{Schwabe2020,Lague2023}, a hybrid framework which evolves the FDM component via the Schrödinger--Poisson equations and the CDM component via an $N$-body solver. Our aim is to test whether a proxy based on initially suppressed matter power also captures the kinematic imprint of the full wave treatment. The comparison includes differences in initial velocity prescriptions as well as subsequent evolution. To do so we compare simulations sharing matched initial suppression but differing in their dynamics, and ask whether approximate mappings such as the Fluctuating Gunn--Peterson Approximation (FGPA) retain sensitivity to these features or whether they are washed out by non-linear evolution and projection into flux space.

\revstart We find that at $f_{\rm A} = 0.1$ the non-linear matter power spectra of the two treatments differ by only $\sim25$ per cent at $k\sim10\,h\,\mathrm{Mpc}^{-1}$ by $z=4$, while their line-of-sight velocity power spectra diverge more strongly at the same scale; the corresponding \LyA flux power spectra differ by over $20$ per cent at $z = 4$, in a regime where linear-theory analyses predict weak sensitivity to the FDM component. This difference includes an inherited initial-velocity offset between the two treatments (Section~\ref{sec:Method}). \revend

The paper is organised as follows: in Section~\ref{sec:Background} we review the theoretical description of FDM, fundamentals of the \LyA forest, and outline our relevant considerations when setting up the simulations. In Section~\ref{sec:Method} we discuss our simulation setup from a cosmic density perturbation to generating the \LyA spectra, and the effects of various parameter choices on the system. Section~\ref{sec:Statistics} examines the results. We conclude in Section~\ref{sec:Conclusion}.

\section{Background and Scope}\label{sec:Background}

\subsection{Ultralight dark matter}

Fuzzy Dark Matter is a hypothetical dark matter candidate composed of axion-like particles (ALPs) with a very small mass and thus an immense de Broglie wavelength. Owing to their immense occupation number in any cosmological structure, the dynamics of FDM in an expanding universe can be captured by a coherent wavefunction that obeys the Schrödinger--Poisson equation, written in a comoving fashion as

\revstart \begin{subequations}\label{eq:FullEOMe}
\begin{align}
i\hbar\partial_t\psi &= \left[-\frac{\hbar^2}{2a^2\maxi}\laplacian + \maxi\Phi\right]\psi,\label{eq:FullSoE}\\
\laplacian\Phi &= \frac{4\pi G}{a} (\abs{\psi}^2 - \langle\rho\rangle), \label{eq:FullSPE}
\end{align}
\end{subequations}
where $a$ is the cosmological scale factor, $t$ is cosmic time, $\vec{\nabla}$ acts on comoving coordinates, $\Phi$ is the gravitational potential, and $\maxi$ is the FDM particle mass. We use a mass-normalised comoving wavefunction, so that $|\psi|^2$ is the comoving FDM mass density and $\langle\rho\rangle$ is its spatial mean. The Poisson equation above describes pure FDM; in the mixed system the particle density also contributes to the source, as specified in Section~\ref{sec:Method}. \revend

The wave-mechanical nature of FDM leaves distinct imprints across 
cosmic time: at high redshifts, the suppression of small-scale power delays the formation of the first collapsed objects and modifies their morphology \citep{Kulkarni2022}, while at intermediate redshifts $2 \lesssim z \lesssim 4$ the FDM velocity field may leave kinematic signatures in the Lyman-$\alpha$ forest that a particle-only proxy does not capture, a possibility we test in this work within an idealised framework.

We adopt the convention of denoting the axion mass in terms of $m_{22} := \maxi/10^{-22}$eV, and will focus on the case where $m_{22} = 0.01$ with an abundance of $\fA := {\Omega_\text{FDM}}/\left({\Omega_\text{FDM} + \Omega_\text{CDM}}\right) = 0.1$. This benchmark is compatible with the current constraints considered in  \citep{Eberhardt2025}.

\subsection{The \LyA forest}

The Lyman-$\alpha$ forest consists of a dense series of absorption features in the spectra of distant quasars, tracing the distribution of neutral hydrogen in the Intergalactic Medium (IGM) across a wide range of redshifts ($2 < z < 6$). These features arise from the \LyA transition of hydrogen which has a rest-frame wavelength of 121.567 nm. As a preeminent probe of the 1D matter power spectra on scales of $1\text{--}10 \, \mathrm{Mpc}/h$, the \LyA forest provides unique constraints on the nature of Dark Matter, particularly in models that predict small-scale power suppression like Fuzzy Dark Matter (FDM) or Mixed Dark Matter (MDM).

With the advent of Stage IV cosmological surveys, such as the Dark Energy Spectroscopic Instrument (DESI, \cite{DESILyADR1}), the statistical uncertainty in the 1D flux power spectrum is reaching the sub-percent level. This observational precision necessitates a corresponding increase in the fidelity of theoretical and numerical models. In this work, we address the challenge of modeling the non-linear evolution of MDM using the \axionyx~code. In this first effort, radiative heating and cooling processes of baryonic matter are not explicitly modeled. Instead of tracking thermal evolution directly, the intergalactic medium (IGM) temperature is prescribed as a function of the total matter density and cosmological redshift in the manner of a Fluctuating Gunn-Peterson Approximation (FGPA,~\citet{Gnedin1998}), an implementation of which we describe in detail in Section \ref{sec:fgpa_pipeline}. By adopting the FGPA, we isolate the gravitational and microphysics effects on the matter distribution against an idealised thermal evolution of the intergalactic medium, allowing us to test the degeneracy between density and velocity suppression in the absence of complex thermodynamics and stellar feedback.

\subsection{Cosmological parameters}

\revstart Table \ref{tab:simulation_parameters} presents key physical parameters used throughout this work, based on \citep{Planck18}. Consistency of parameters and unit systems across the various software packages we employ is extensively verified. \revend

\begin{table}
    \centering
    \begin{tabularx}{\columnwidth}{X l}
        \toprule
        \textbf{Parameter} & \textbf{Value} \\
        \midrule
        \(h\) & 0.675 \\
        \(\Omega_\text{b}\)\textsuperscript{a} & 0.0487 \\
        \(\Omega_\text{DM}\) & 0.2613 \\
        \(m_{22}\) & 0.01 \\
        \(f_\text{A}\) & 0.1 \\
        \(n_s\) & 0.96 \\
        \(\sigma_8(z = 0)\) & 0.811 \\
        \(z_\text{Start}\) & 120 \\
        \(z_{\text{End}}\) & 2 \\
        \addlinespace
        Resolution & \(1024^3\) \\
        Box Length & 30 Mpc/\(h\) \\
        \bottomrule
    \end{tabularx}
    \caption{Parameters used for the initialisation and simulation of the MDM suite.
    \textsuperscript{a}Used in the IC generation and FGPA pipeline only; not in the simulation itself.}
    \label{tab:simulation_parameters}
\end{table}

\section{Method}\label{sec:Method}

\revstart This work compares three simulations sharing identical background cosmologies and initial random phases, differing only in the dark matter treatment: \revend

\begin{itemize}
    \item \textbf{$\Lambda$CDM}: reference box with no FDM component, evolved with an $N$-body solver.
    \item \textbf{MDM Full}: the same random phases and total-matter suppression as MDM $N$-body, but with the axion component represented as a wavefunction and evolved with full Schrödinger--Poisson dynamics.
    \item \textbf{MDM $N$-body}: FDM effects entering \emph{only} through the suppressed initial matter power spectrum, thereafter evolved with the same $N$-body solver.
\end{itemize}

\revstart We compare a two-component MDM realisation evolved with full Schrödinger–Poisson dynamics to a single-species $N$-body realisation constructed from the same random phases and matched total-matter transfer function. This comparison holds the initial total-matter suppression fixed, while allowing the initial velocity prescription, subsequent component dynamics and phase-space representation to differ. \revend

\revstart The comparison tests what is missed by a particle-only proxy that encodes FDM through suppressed initial density power. It is related to the pure-FDM comparison of \citet{Li2019}, who evolved the same FDM initial conditions with wave and $N$-body dynamics. In our mixed model, however, the initial velocity prescriptions also differ: the full calculation uses species-specific velocity transfers, whereas the proxy uses CDM-like velocities (Section~3.1). The measured difference therefore combines the inherited velocity offset with the subsequent dynamical evolution, rather than isolating the latter alone. \revend

While \axionyx~ supports Adaptive Mesh Refinement (AMR), the simulations in this work were performed on a fixed $1024^3$ root grid. This ensures uniform spatial resolution across the entire volume, which is critical for the unbiased extraction of the flux power spectra. Furthermore, by avoiding multi-level mesh transitions, we eliminate potential interpolation artifacts that could otherwise interfere with the subtle small-scale suppression signal of the mixed dark matter models.

\subsection{Initial condition generation}
\begin{figure}
    \centering
    \includegraphics[width=\linewidth]{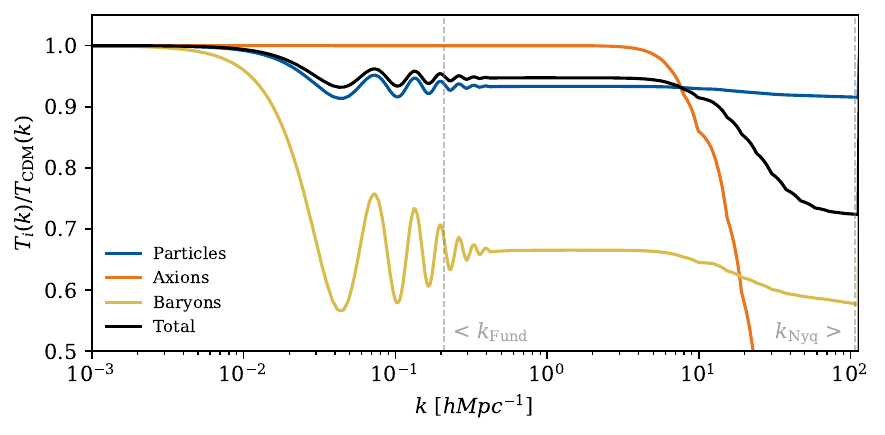}
    \caption{\color{revcolor} Initial density transfer functions for the FDM and CDM components for $m_{22} = 0.01$ and $\fA = 0.1$. The curves are normalised against the CDM transfer function $T_\text{CDM}(k)$. The baryonic transfer function is not plotted separately, but the BAO signatures are visible in the total matter and particles curves. The Fundamental and Nyquist wavenumbers of simulation box are marked on the $x$-axis.
    }
    \label{fig:axionCAMB}
\end{figure}

The linear transfer functions were computed using \texttt{AxiECAMB}
\citep{AxiECAMB, AxionCAMB22}. For the $\Lambda$CDM reference run, we set the axion fraction to a negligible value, such that all three boxes are produced by a single, consistently-configured code. To represent the three-component universe (CDM, baryons, and axions) within our two-component simulation framework, we group the baryonic and CDM components into an effective ``particle'' species. Given that baryons are approximately collisionless and tightly coupled to the CDM potential at $z \gtrsim 100$, we define an effective particle transfer function as a mass-weighted sum:
\begin{equation}
    T_{\text{P}}(k) = \frac{\Omega_c T_{\text{c}}(k) + \Omega_b T_{\text{b}}(k)}
    {\Omega_c + \Omega_b}.
\end{equation}
This approximation neglects small-scale baryon pressure and relative velocity
offsets, which are small on the IGM scales considered here. The resulting transfer functions for the $10$ per cent FDM case are shown in Fig.~\ref{fig:axionCAMB}.

\revstart The primordial scalar amplitude $A_s$ is not taken directly from
Planck but is instead tuned for each dark matter
configuration, such that the linear total-matter $\sigma_8$ extrapolated to
$z=0$ equals $0.811$. This ensures the three boxes share a common
large-scale normalisation despite their differing small-scale suppression. \revend

The simulations are initialised at $z=120$ using second-order Lagrangian Perturbation Theory (2LPT), as implemented in the public \texttt{MUSIC} code
\citep{MUSIC}. Following the methodology of \citet{Lague2023}, all species are
seeded from a common set of random phases and modulated by the species-specific transfer functions. This phase-matching ensures the resulting density fields remain physically correlated across species, as expected for fluctuations originating from a single inflationary source.

\revstart The axion component is realised as a grid-based wavefunction $\psi=\sqrt{\rho}\,e^{i\theta}$ on the grid, with the amplitude set by the axion density transfer function $T_{\text{axion}}(k)$ and the phase by the axion velocity transfer function, $\mathbf v=(\hbar/am)\nabla\theta$. Both are computed in the Newtonian gauge by a modified version of \texttt{AxiECAMB}. For the particle sector, $N$-body particles of uniform mass are positioned on a regular grid and displaced according to the displacement field derived from $T_{\text{P}}(k)$, with velocities from the corresponding mass-weighted Newtonian-gauge velocity transfer function of CDM and baryons. \revend

\revstart The suppressed MDM $N$-body box requires a single particle species to carry the full matter budget, and its initial conditions are constructed accordingly. The displacement field is generated from the total matter transfer function $T_{\rm tot}(k)$, while particle velocities are assigned using the Newtonian-gauge CDM velocity transfer function. This combination of suppressed displacements, CDM-like velocities matches the convention used in observational FDM \LyA constraints from $N$-body simulations with suppressed initial conditions \citep{RogersPeiris2021}. \revend

\revstart  \revend

\subsection{Advancing the simulation}

\revstart We evolve the MDM system using the public \axionyx~ code, a hybrid cosmological simulation code developed from \texttt{Nyx} and the performant \texttt{Amrex} code family \citep{Almgren2013,Almgren2019}.  \revend

The axion wavefunction $\psi$ is evolved on the periodic root grid using a high-order split-operator pseudo-spectral solver. By performing the kinetic drift in Fourier space, we achieve spectral accuracy for the spatial Laplacian, which is critical for correctly capturing the ``quantum pressure'' term, $\mathcal{Q} \propto \nabla^2 \sqrt{\rho} / \sqrt{\rho}$. 

\revstart Time evolution is achieved using a 6th-order symplectic integrator. This multi-stage split-operator scheme (utilizing a kick-drift-kick sequence) ensures high-order conservation of the Hamiltonian and phase accuracy over cosmological timescales. Combined gravity is solved via a multi-level Poisson solver:
\begin{equation}
    \nabla^2 \Phi = \frac{4\pi G}{a} (\rho_{\mathrm{part}} + |\psi|^2 - \bar{\rho}),
\end{equation}
where $\rho_{\mathrm{part}}$ is the comoving particle-mesh (PM) density of the effective CDM-plus-baryon particle species, projected to the grid via a Cloud-in-Cell (CIC) kernel, and $\bar{\rho}$ is the mean total comoving matter density.  \revend

A significant challenge for MDM simulations is the disparity in characteristic timescales between the heavy CDM particles and the light axion field. In our hybrid scheme, the global simulation timestep $\Delta t$ is determined by the most restrictive of several physical constraints, evaluated on the fly. For the axion component, the stability of the Schrödinger evolution is governed by the phase frequency of the wavefunction. On the root grid, this manifests as a combination of the free-particle dispersion and the local gravitational potential:
\begin{equation}
    \Delta t_{\mathrm{axion}} = \min \left( \eta \frac{a^2 \Delta x^2}{6 (\hbar/m)}, \frac{\hbar/m}{|\Phi_{\mathrm{max}} - \Phi_{\mathrm{min}}|} \right),
\end{equation}
where $\Delta x$ is the grid spacing, and $\eta$ is a user-specified safety factor. The first term represents the von Neumann stability criterion for the kinetic operator, while the second term ensures that the phase evolution due to the gravitational potential $\Phi$ is well-sampled.

On the other hand, the CDM sector must satisfy the standard cosmological Courant-Friedrichs-Lewy (CFL) condition to ensure that particles do not traverse more than a fraction of a cell width per step. By dynamically evaluating these criteria at each step, \axionyx ~ maintains phase coherence in the wave sector while accurately tracking the non-linear clustering of the particle sector.

\subsection{Ly$\alpha$ spectra via FGPA}
\label{sec:fgpa_pipeline}

\revstart We generate mock Ly$\alpha$ forest transmission skewers from our dark-matter-only simulations using a bespoke implementation of the FGPA forward model. This differs from the local Gunn–Peterson treatment adopted by \citet{Li2019}, in which the optical depth was constructed directly from a smoothed density field. Here the neutral fraction and thermal width are computed explicitly within the FGPA prescription, and the simulated line-of-sight peculiar velocity enters through the redshift-space mapping described below. \revend

\revstart From each snapshot we extract one-dimensional skewers along each of the three Cartesian axes. Every skewer spans the full $1024$ cells of the box along its line-of-sight direction. For memory management, the transverse plane is divided into $32$ tiles of $256\times128$ cells; an edge margin of $8$ cells is applied to each tile to suppress transverse filtering artefacts. Sampling every fourth remaining transverse grid point gives $60\times28$ skewers per tile, or $53{,}760$ per line-of-sight direction and $161{,}280$ in total. Within each tile the transverse separation is $\simeq0.17$ cMpc. This exceeds the adopted filtering scale $\lambda_{\rm F}$ (Eq.~\ref{eq:lambdaF}), but does not imply statistical independence of neighbouring sightlines. The Fourier transform covers the complete sightline; no longitudinal partitioning is performed. The output-pixel test in Section~\ref{sec:Statistics} addresses spectral sampling while holding this skewer selection fixed. \revend

For each skewer we compute the LOS peculiar velocity $v_{\parallel}(x)$, the neutral hydrogen number density $n_{\rm HI}(x)$, and the thermal Doppler parameter $b(x)$.

\subsubsection{Matter density and baryon tracing.}
\revstart The FGPA rests on the assumption that, above the pressure filtering scale, the gas is unsupported against gravity and therefore traces the total matter distribution with a fixed mass ratio. Below that scale, pressure decouples the two. We accordingly begin from the total simulated dark matter density $\rho_{\rm m}(\mathbf{x})$, which comprises both the $N$-body particle component and, in the mixed case, the axion field $|\psi|^2$, and set the physical baryon density to
\begin{equation}
\rho_{\rm b}(\mathbf{x}) \equiv \frac{\Omega_{\rm b}}{\Omega_{\rm m}}\,\rho_{\rm m}(\mathbf{x})\,(1+z)^3,
\end{equation} \revend

\revstart This neglects any scale-dependent bias between gas and dark matter above $\lambda_{\rm F}$, which is a good approximation in the moderate-overdensity IGM ($\delta \sim 1$--$10$) to which the \LyA forest is most sensitive, but degrades in collapsed structures where shock heating and radiative cooling become relevant. \revend

\revstart We apply a phenomenological baryonic pressure smoothing to the density and velocity fields using a Gaussian filter with width
\begin{equation}\label{eq:lambdaF}
    \lambda_\mathrm{F} \approx 0.2\,\mathrm{Mpc}
    \left(\frac{1}{1+z}\right)^{1/2},
\end{equation}
following \citet{Gnedin1998}. We note that this is strictly justified only in linear theory, where pressure smoothing in the density and velocity fields is tightly coupled; following non-linear hydrodynamic response at late times the two need not share a common filtering scale. Absent a hydrodynamic treatment we have no self-consistent way to separate them, and we adopt a single kernel as a phenomenological prescription. We quantify the sensitivity of our conclusions to the velocity field in
Section~\ref{sec:Statistics}. We also verified that filtering the momentum $\rho\vec{v}$ and dividing, rather than filtering $\vec{v}$ directly, changes the suppression signal negligibly. \revend

\subsubsection{Axion/FDM velocity field and MDM momentum weighting.}

The FDM velocity field $\vec{v}$ is given by the gradient of the complex phase $\theta$:
\begin{equation}
    \vec{v} = \frac{\hbar}{a \maxi} \vec{\nabla} \theta ,
\end{equation}
where $\vec{\nabla}$ represents the gradient in comoving coordinates. Due to the wraparound nature of the phase, performing explicit finite difference on $\theta$ directly is numerically unstable. We thus recall that 
\begin{equation}\label{eq:phasegrad}
    \vec{\nabla}\theta = \frac{\Im\left(\psi^*\vec{\nabla}\psi \right)}{\rho} = \frac{\Re(\psi)\vec{\nabla}\Im(\psi) - \Im(\psi)\vec{\nabla}\Re(\psi)}{\rho} ,
\end{equation}
where $\rho = |\psi|^2$ is the field density. This formulation allows the velocity to be computed using standard finite-difference derivatives of the real and imaginary components of $\psi$ extracted from simulations, which are well-behaved across the domain. \revstart We evaluate the line-of-sight derivatives with second-order periodic central differences. Replacing them with a spectral derivative changes the $z=4$ line-of-sight velocity power of the full box by $\lesssim4$ per cent ($\sim2$ per cent at $k=10\,h\,\mathrm{Mpc}^{-1}$), raising it slightly, so the periodic scheme marginally overstates the divergence from the $N$-body treatment. \revend Vortex cores of the wavefunction produce localised velocity spikes near the $\rho\to0$ nodes in Eq.~\ref{eq:phasegrad}; a density floor guards this division against them. For the fiducial density-weighted construction this floor is never engaged; in the global-fraction ablation of Section~\ref{sec:Statistics}, which requires a standalone axion velocity, it activates for a negligible fraction of cells ($\sim4\times10^{-5}$ per cent), with no measurable impact on the resulting power spectrum.

\revstart We then construct a total LOS velocity by density-weighting the CDM and FDM components,
\begin{equation}
v_{\parallel}(\mathbf{x})
=
\left[1-f_{\rm a}(\mathbf{x})\right]v^{\rm (c)}_{\parallel}(\mathbf{x})
+
f_{\rm a}(\mathbf{x})v^{\rm (a)}_{\parallel}(\mathbf{x}),
\label{eq:vlos_composite}\end{equation}
with the local FDM fraction of the represented total matter
\begin{equation}\label{eq:LocalSpeedMix}
f_{\rm a}(\mathbf{x})=\frac{\rho_{\rm a}(\mathbf{x})}{\rho_{\rm a}(\mathbf{x})+\rho_{\rm part}(\mathbf{x})}.
\end{equation}
Here $v^{\rm(c)}_{\parallel}$ denotes the velocity of the effective particle species, whose density $\rho_{\rm part}$ includes the CDM and baryonic budgets. Thus $f_{\rm a}(\mathbf{x})$ has cosmic mean $\Omega_{\rm FDM}/\Omega_{\rm m}\simeq0.084$, whereas $f_{\rm A}=0.1$ is defined relative to dark matter alone. This local weighting allows the axion fraction to fluctuate spatially. In Section \ref{sec:Statistics}, we consider two further velocity weighting schemes: using a single global axion fraction of total matter, $\Omega_{\rm FDM}/\Omega_{\rm m}\simeq0.084$, and using CDM speed information only. The suppression in \LyA spectra is robust across the definitions, but is the most pronounced when we use the localised fraction. \revend

\revstart \subsubsection{FGPA thermodynamics and ionisation equilibrium}
We model the low-density intergalactic medium with a power-law temperature--density relation,
\begin{equation}
\mathcal{T}(\mathbf{x}) = \mathcal{T}_0\,\Delta(\mathbf{x})^{\gamma-1},
\qquad 
\Delta(\mathbf{x}) \equiv \frac{\rho_{\rm b}(\mathbf{x})}{\bar\rho_{\rm b}(z)},
\end{equation}
with fiducial parameters $\mathcal{T}_0\simeq 10^4\,{\rm K}$ and $\gamma\simeq 1.55$ following common choices in the literature (e.g.\ \citep{Lukic2014}). In the implementation, $\Delta$ is clipped to $[10^{-8},100]$ when evaluating this temperature relation, and a temperature floor of $1000$ K is applied. A numerical Doppler-width floor of $1\,\mathrm{km\,s^{-1}}$ is also imposed, but is inactive under this temperature floor. 
Given $\mathcal{T}(\mathbf{x})$, we evaluate the Case-A recombination coefficient using a standard power-law approximation,
\begin{equation}
\alpha_{\rm A}(\mathcal{T}) \simeq 4.2\times 10^{-13}\left(\frac{\mathcal{T}}{10^4\,{\rm K}}\right)^{-0.7}\,{\rm cm^3\,s^{-1}},
\end{equation}
and assume photoionisation equilibrium with a spatially uniform hydrogen photoionisation rate $\Gamma_{\rm HI}$:
\begin{equation}
n_{\rm HI}(\mathbf{x})
=
\frac{n_{\rm H}(\mathbf{x})\,n_{\rm e}(\mathbf{x})\,\alpha_{\rm A}[\mathcal{T}(\mathbf{x})]}{\Gamma_{\rm HI}},
\end{equation}
with $n_{\rm H}=X_{\rm H}\rho_{\rm b}/m_{\rm p}$ and $n_{\rm e}\simeq 1.08\,n_{\rm H}$ to account for singly ionised helium. Thermal broadening is included via the Doppler parameter
\begin{equation}
b(\mathbf{x})=\sqrt{\frac{2k_{\rm B}\mathcal{T}(\mathbf{x})}{m_{\rm p}}}.
\end{equation} \revend

\revstart The FGPA as implemented here does not capture fluctuations in $n_{\rm HI}$ that arise independently of density. In a real intergalactic medium the temperature field carries the imprint of inhomogeneous reionisation histories and of local radiative processes, so that the neutral fraction is correlated with, but not a deterministic function of, the gas density. A treatment that captured this would require radiative transfer and a thermal history we do not attempt to model here. \revend

\revstart We report flux-power ratios between simulations sharing an identical FGPA prescription. This can reduce sensitivity to some shared modelling choices, but the nonlinear mapping acts on different density and velocity fields, so cancellation of systematic errors is not guaranteed. We test one aspect of the thermal sensitivity by varying $\mathcal{T}_0$, which changes both the recombination coefficient and thermal broadening. Within the adopted FGPA prescription, the MDM Full-to-$\Lambda$CDM flux-power ratio is relatively insensitive to the tested temperature normalisations (Fig.~\ref{fig:Result1_Thermals}). This test does not establish robustness to thermal-history uncertainties or nonlinear baryonic response. \revend

\subsubsection{Redshift-space mapping and transmitted flux.}
Along each skewer we map comoving position $x$ to LOS velocity coordinate $u$ by combining the Hubble flow with peculiar velocity,
\begin{equation}
u(x) = \frac{H(z)}{1+z}\,x + v_{\parallel}(x),
\end{equation}
and enforce domain periodicity by wrapping $u$ into $[0,V_{\rm max})$ with
\begin{equation}
V_{\rm max} \equiv \frac{H(z)}{1+z}\,L_{\rm box}.
\end{equation}

\revstart We compute the Ly$\alpha$ optical depth $\tau(u)$ by summing contributions from all real-space cells using a thermally broadened line profile. Using the Ly$\alpha$ oscillator strength $f_{12}$ and rest wavelength $\lambda_0$, the cross-section prefactor is
\begin{equation}
\sigma_0 = \frac{e^2 f_{12}\lambda_0}{4\epsilon_0 m_e c}.
\end{equation}
For each output velocity pixel $u_i$ we approximate
\begin{equation}
\tau(u_i) \approx 
\sum_j
\frac{n_{\rm HI}(x_j)\,\sigma_0}{H(z)}
\,
\phi\!\left(\frac{u_i-u(x_j)}{b(x_j)}\right)\,
\Delta u,
\end{equation}
\begin{equation}
\phi(y)=\frac{1}{\sqrt{\pi}\,b}\exp(-y^2),
\end{equation}
where $\Delta u \simeq [H(z)/(1+z)]\Delta x$ is the velocity spacing associated with the uniform real-space grid. We then form the transmitted flux
\begin{equation}
F(u) = \exp[-\tau(u)].
\end{equation}
Because $\tau$ is evaluated directly at every pixel $u_i$ of a uniform periodic velocity grid, no interpolation or rebinning step is needed between the optical depth and the flux. We further introduce an additional amplitude parameter $A$ such that $F=\exp[-A\tau]$ and calibrate $A$ to match a fiducial mean flux $\bar{F}(z)$. To this end we adopt the fit of \citet{Onorbe2017},
\begin{equation}\label{eq:meandepth}
    \tau_{\rm HI}(z) = 0.00126 \exp\!\left(3.294\sqrt{z}\right)~.
\end{equation} \revend

\subsubsection{Neutral-hydrogen weighting}\label{sec:NHIWeighting}
\revstart The composite velocity of Eq.~\ref{eq:vlos_composite} weights the effective particle and FDM components by their local shares of total matter (Eq.~\ref{eq:LocalSpeedMix}), not by the neutral hydrogen fraction. The velocity power spectrum we report is therefore a statement about the dark matter kinematics, whereas the quantity that sets the forest is the neutral-hydrogen-weighted velocity. The two enter differently. In the flux calculation, neutral-hydrogen weighting is already present at the level of the optical depth: each cell contributes to $\tau(u_i)$ in proportion to $n_{\rm HI}(x_j)$, placed at its own redshift-space position $u(x_j)$. Under the adopted temperature--density relation and photoionisation equilibrium, $n_{\rm HI}\propto\Delta^{2-0.7(\gamma-1)}\approx\Delta^{1.6}$ for $\gamma=1.55$ where the temperature clipping and floor are inactive, so absorption is biased towards overdense gas, where the CDM share of the composite velocity is larger. The same holds for the temperature: the Doppler parameter $b(x_j)$ of each cell enters its own line profile in the sum for $\tau(u_i)$, weighted by that cell's $n_{\rm HI}$, so the effective thermal broadening is neutral-hydrogen-weighted by construction. Likewise, no velocity is averaged over a pixel: each cell's velocity displaces its own absorption to $u(x_j)$, so an unweighted mean over the most dynamically active regions does not enter the flux. The weighting questions that remain concern the diagnostic velocity power spectrum and the temperature--density relation itself. \revend

\revstart The velocity power spectra of Figs.~\ref{fig:SpeedPS} and \ref{fig:FullVsNbody}, and the $\sim44$ per cent quoted above, are the unweighted and unfiltered line-of-sight velocity power of the composite field. They are not the velocity that enters the flux calculation, which is smoothed with the Gaussian filter of Eq.~\ref{eq:lambdaF}. Applying that filter before computing the power spectrum raises the $z=4$ full-to-$N$-body ratio from $0.44$ to $0.58$ at $k=10\,h\,\mathrm{Mpc}^{-1}$ and from $0.23$ to $0.50$ at $k=20\,h\,\mathrm{Mpc}^{-1}$, so the velocity divergence is smaller than the unfiltered diagnostic suggests, but at $k\sim10\,h\,\mathrm{Mpc}^{-1}$ it remains larger than that of the matter power ($0.75$). A neutral-hydrogen-weighted version of the filtered diagnostic, with each cell weighted by $n_{\rm HI}$, gives $0.72$ and $0.67$ at the same wavenumbers, so weighting reduces the divergence further but does not remove it. We regard this weighted result as preliminary and do not use it in the figures: the weighted line-of-sight mean velocity need not vanish along a skewer, which can bias the large-scale modes of the weighted spectrum independently of the dark matter treatment, and a fully specified estimator is left to future work. This is distinct from the per-cell neutral-hydrogen weighting already present in the optical-depth calculation. A self-consistent comparison with the absorbing gas also requires hydrodynamical simulations in which $n_{\rm HI}$, $\mathcal{T}$ and the baryonic velocity respond to the pressure and thermal history of the gas. \revend

\section{Results and statistics}\label{sec:Statistics}

\begin{figure}
    \centering
    \includegraphics[width=\linewidth]{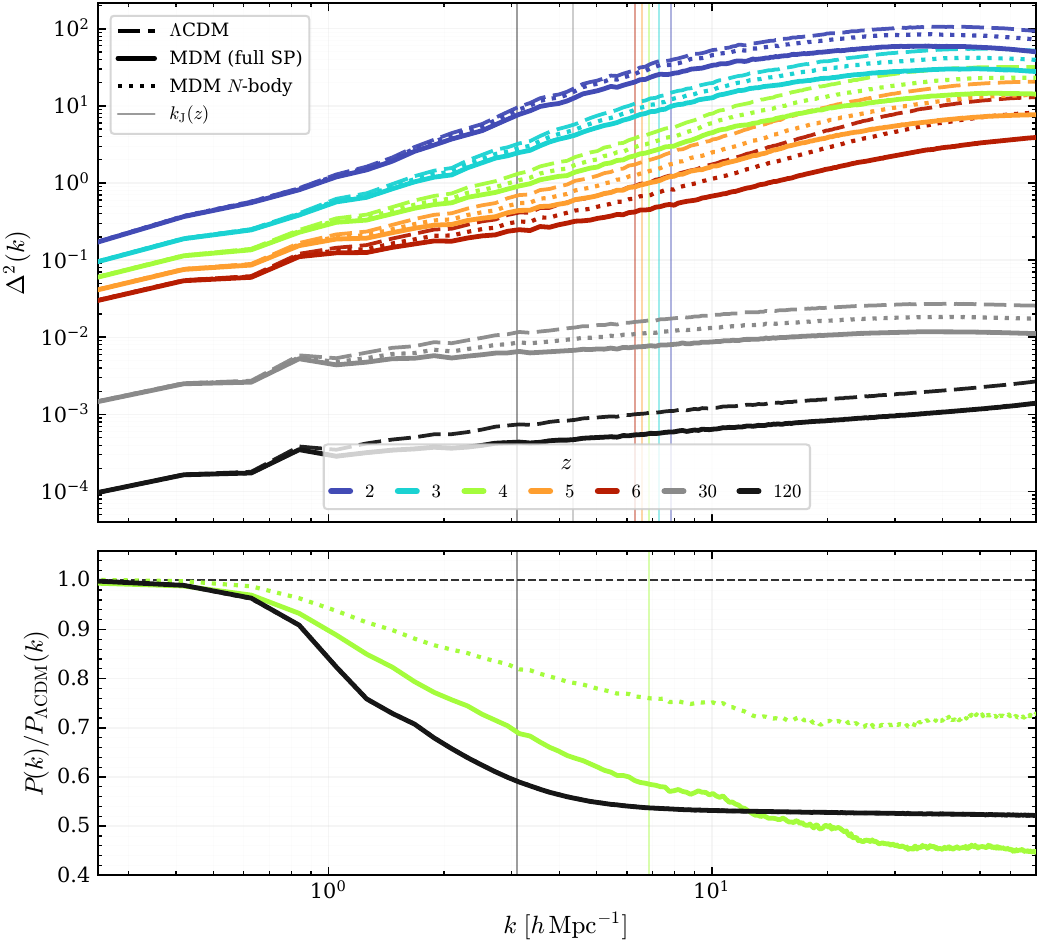}
\caption{%
\textbf{Top:} Dimensionless matter power spectra, $\Delta^2(k)$, at simulation snapshots for
$z=2$, 3, 4, 5, 6, 30 and 120 (colour). Dashed, solid, and dash-dotted curves show the $\Lambda$CDM baseline, the full MDM calculation, and the MDM $N$-body-only approximation, respectively. Vertical lines mark the corresponding Jeans wavenumbers, $k_\mathrm{J}(z)$, colour-matched to each redshift. \newline \textbf{Bottom:} Ratio of the MDM power spectra to the $\Lambda$CDM power spectrum, isolated at $z = 4$ and $z = 120$. The full MDM calculation exhibits stronger suppression than the $N$-body-only approximation.}
    \label{fig:NLGrowth}
\end{figure}

\begin{figure*}
    \centering
    \includegraphics[width=0.61\linewidth]{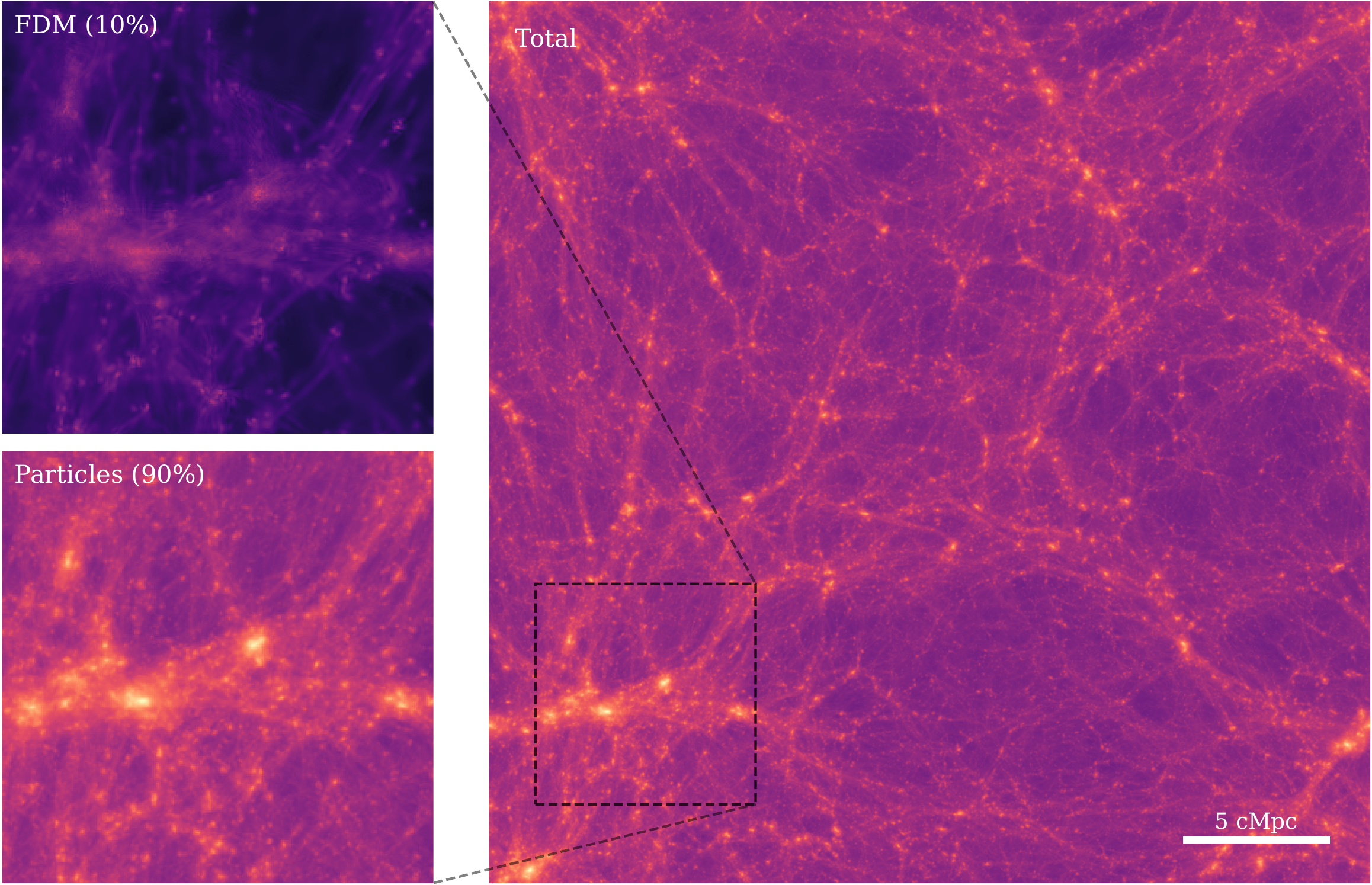}
    \caption{Projected density fields at $z = 2$ within the $30\,\text{Mpc}/h$ simulation box. The \textbf{left panels} show a zoomed-in subregion (width $\approx 8\,\text{cMpc}$ around a dense region) highlighting the distinct morphologies of the particle (CDM) and FDM components. The \textbf{right panel} displays the total projected density across the full simulation volume, with the location of the zoomed subregion indicated on it. While the CDM follows the standard cosmic web morphology, the FDM component exhibits characteristic wave-mechanical interference patterns and smoothed small-scale structures that contribute to the total mass distribution shown on the right. All plots use the same color scale.}
    \label{fig:Nyx_B}
\end{figure*}

\revstart \begin{figure}
    \centering
    \includegraphics[width=\linewidth]{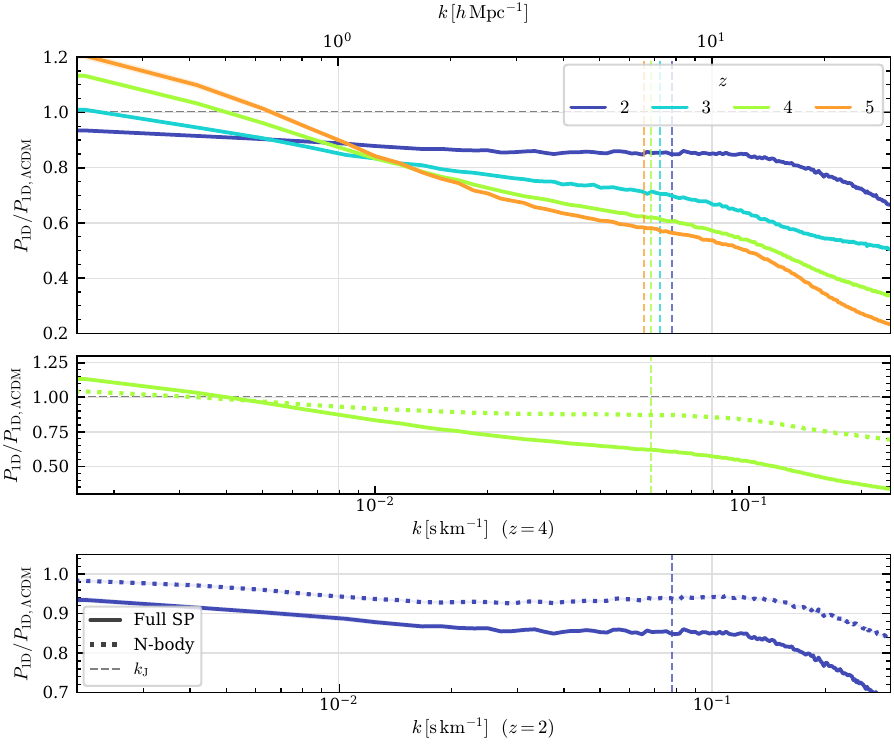}
    \caption{\color{revcolor} \textbf{Top:} Ratio of the \LyA 1D flux power spectrum in the full Schrödinger--Poisson MDM treatment to the corresponding $\Lambda$CDM result, at $z=5$, 4, 3 and 2. \textbf{Middle, bottom:} the same ratio isolated at $z = 4$ and $z = 2$ respectively, now comparing the full Schrödinger--Poisson treatment (solid) against the $N$-body approximation with matched initial total-matter density power and random phases (dotted), to highlight where the two diverge at fixed redshift. Mean fluxes are calibrated to the same fiducial value at each redshift. The horizontal dashed line marks equality with $\Lambda$CDM; vertical dashed lines mark the axion Jeans wavenumber $k_J(z)$, colour-matched to each redshift. The top axis of the uppermost panel gives $k$ in $h\,\mathrm{Mpc}^{-1}$ and is shared by all three panels; the bottom axes of the middle and lower panels give the corresponding $k$ in $\mathrm{s/km}$ at their own redshift.}
    \label{fig:Result1}
\end{figure} \revend

\begin{figure}
    \centering
    \includegraphics[width=\linewidth]{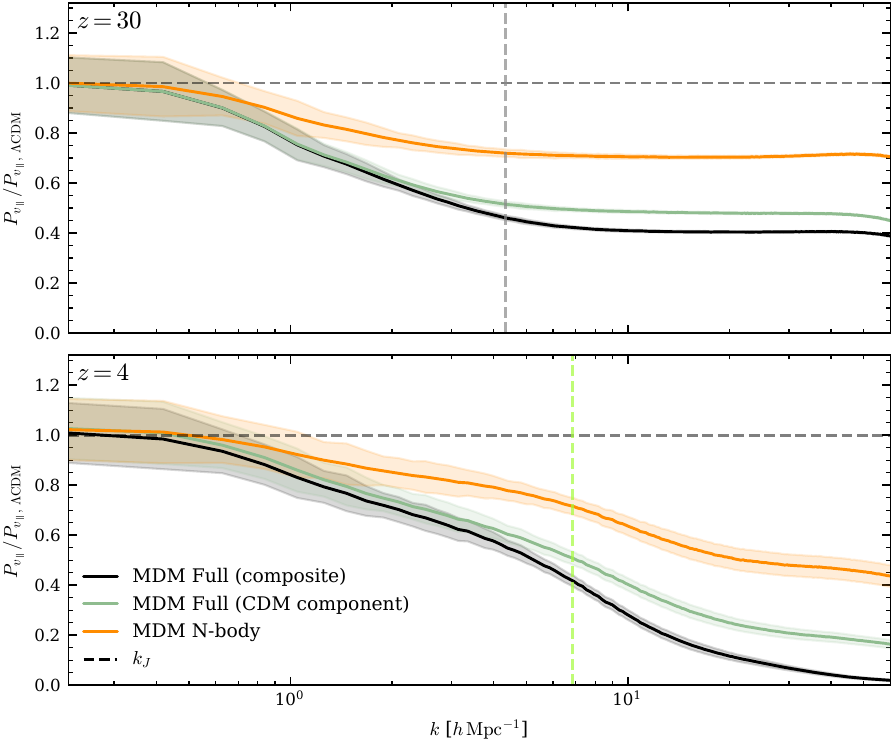}
    \caption{Line-of-sight velocity power spectra at $z=30$ and $z = 4$, shown as ratios to the $\Lambda$CDM baseline. The CDM component in the MDM Full simulation is plotted separately in light green. Velocity power is computed from the unfiltered composite field (Section~\ref{sec:NHIWeighting}). At $z=30$, which is still in the linear regime, the $N$-body box retains $\sim70$ per cent of the $\Lambda$CDM power on small scales and the full composite $\sim40$ per cent; this offset is inherited from the initial conditions (Section~\ref{sec:Method}). By $z = 4$ the gap has widened: the $N$-body approximation retains $\sim 40$ per cent of the $\Lambda$CDM velocity power on small scales, while the full simulation box with full Schrödinger–Poisson treatment drives it to near zero beyond the Jeans scale. We have not isolated the physical origin of this suppression.}
    \label{fig:SpeedPS}
\end{figure}

\begin{figure*}
    \centering
    \includegraphics[width = 0.9\textwidth]{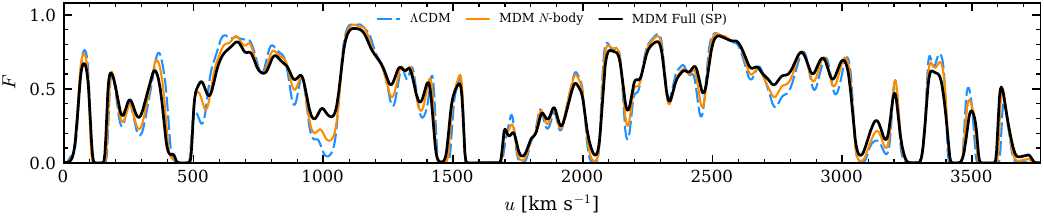}
   \caption{The transmitted flux along a single line of sight at $z = 4$, for the three simulations. The sightline is the same in all three boxes, which share initial random phases, and was selected as that whose mean flux lies closest to the sample median. Note that the full Schrödinger--Poisson treatment (black) shows visibly fewer fine structures within them than either the $N$-body approximation (orange) or $\Lambda$CDM (blue).}
    \label{fig:Skewer}
\end{figure*}

\revstart 
\begin{figure}
    \centering
    \includegraphics[width=\linewidth]{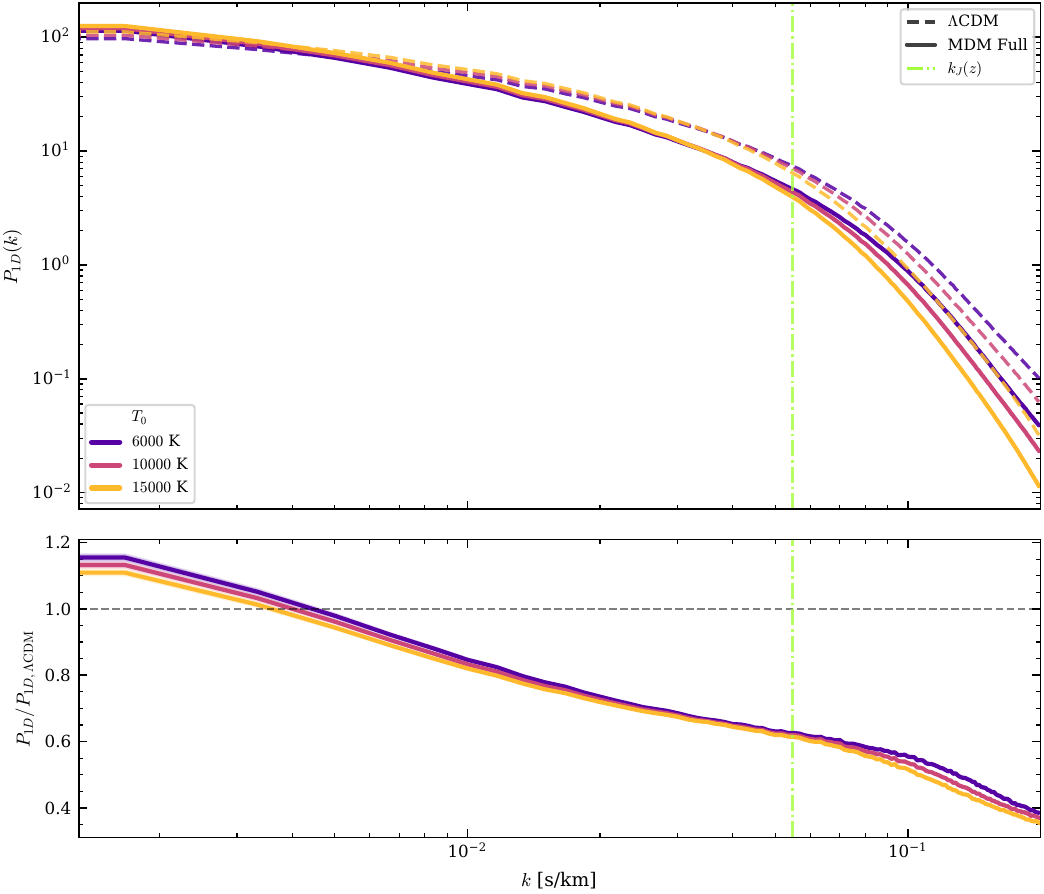}
    \caption{\color{revcolor} \textbf{Top:} \LyA\ transmission power spectrum at $z=4$ in $\Lambda$CDM (dashed) and the full Schr\"odinger--Poisson MDM treatment (solid), for $\mathcal{T}_0 = 6\,000$, $10\,000$, and $15\,000$~K (colour). The vertical dash-dotted line marks the axion Jeans wavenumber $k_J(z=4)$. \textbf{Bottom:} Ratio of the MDM to $\Lambda$CDM power spectra shown above. Within the adopted FGPA prescription, the MDM Full-to-$\Lambda$CDM ratio is relatively insensitive to the tested temperature normalisations, despite the temperature dependence of the absolute spectra. This does not establish robustness to thermal-history uncertainties. For each dark matter model and value of $\mathcal{T}_0$, the mean flux is independently calibrated to match the fiducial value of Eq.~\ref{eq:meandepth}.}
    \label{fig:Result1_Thermals}
\end{figure} \revend

\begin{figure}
    \centering
    \includegraphics[width=\linewidth]{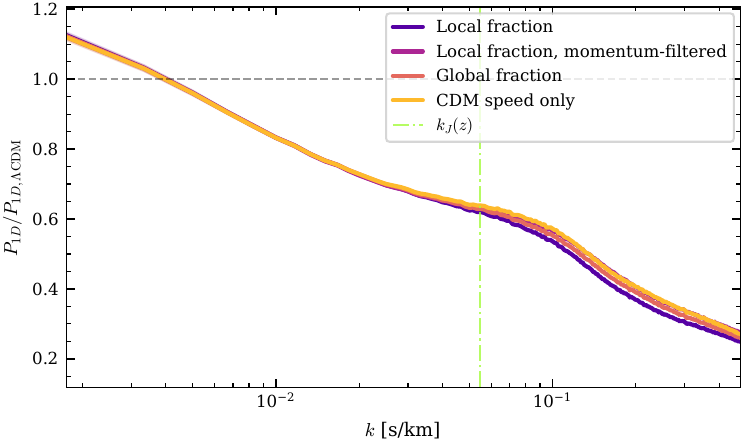}
    \caption{The $z = 4$ \LyA transmission power spectrum compared with the corresponding LCDM result as the strategy of local velocity estimator switches between local fraction (Equation \ref{eq:vlos_composite}), CDM speed only, the global axion fraction of total matter, $\Omega_{\rm FDM}/\Omega_{\rm m}\simeq0.084$, or the local-fraction estimator with the momentum $\rho\vec{v}$ filtered instead of $\vec{v}$ (Section~\ref{sec:Method}). Local fraction yields the largest suppression, as it biases the most towards producing a smoother velocity field.}
    \label{fig:Result1_Weights} 
\end{figure}

The nonlinear matter-power evolution for all three simulation boxes is shown in Figure~\ref{fig:NLGrowth}, and a projected density map of the MDM Full simulation at $z = 2$ is shown in Figure~\ref{fig:Nyx_B}. The matter power spectra are computed with a spectral estimator built on \texttt{AMReX} and \texttt{FFTW}, with the particle component deposited onto the same regular grid on which the FDM field is defined using a cloud-in-cell kernel and subsequently deconvolved. At $z=120$, both mixed-DM runs exhibit the identical suppression on small scales inherited from the FDM transfer function, as expected. By $z = 4$, when nonlinear structure growth is well underway, the matter power spectra of the two mixed-DM treatments diverge by $\sim25$ per cent at $k\sim10\,h\,\mathrm{Mpc}^{-1}$ — smaller than the divergence found below in the velocity field, but not negligible on its own.

\revstart Figure~\ref{fig:Nyx_B} supports the interpretation that, at $f_A=0.1$, the highest-density regions (halos and filament cores) are shaped predominantly by the CDM component, with the FDM distribution largely tracing the same large-scale structures. This helps explain why the matter power spectra, despite starting from identical initial total-matter density power at $z=120$, diverge only modestly by $z=4$ rather than tracking the larger divergence seen in the velocity field. However, the \LyA forest is most sensitive to the extended distribution of neutral hydrogen that comprises the moderate-overdensity IGM ($\delta \sim 1-10$), where the two mixed-DM treatments need not remain degenerate. In this regime, wave-mechanical effects can imprint differences not only in the density structure but also in the coherent LOS velocity field, both of which enter nonlinearly into the FGPA mapping. \revend

The dashed vertical lines in Figs.~\ref{fig:NLGrowth} and \ref{fig:Result1} mark
the effective axion Jeans wavenumber, above which quantum pressure becomes
dynamically significant:
\begin{equation}
    k_{\rm J} = \frac{66.5}{h} \, (1+z)^{-1/4}
    \left( \frac{\Omega_{\rm m} h^2}{0.12} \right)^{1/4}
    m_{22}^{1/2}
    \; \; h\,\mathrm{Mpc}^{-1} .
    \label{eq:kjeans}
\end{equation}

\revstart Our benchmark $m_{22}=0.01$ is 100 times lighter than the commonly used fiducial ULDM mass $m_{22}=1$. At fixed redshift and background cosmology, the scaling $k_J\propto m_{22}^{1/2}$ gives a Jeans wavenumber smaller by a factor of 10, or a Jeans length larger by a factor of 10. This places the characteristic scale of wave effects at larger spatial scales than for that fiducial mass. It does not by itself determine the amplitude of the flux-power difference, which also depends on the FDM fraction, velocity prescription and baryonic mapping. \revend

\revstart The generated Ly$\alpha$ forests are flux-calibrated so that, at each redshift, the mean flux is consistent with Equation \ref{eq:meandepth}. Fig.~\ref{fig:Result1} shows the ratio of MDM Ly$\alpha$ forest power spectra to their respective $\Lambda$CDM baselines across $z = 2$, 3, 4 and 5, with the $z = 2$ and $z = 4$ cases shown separately in the lower panels; these bracket the range over which the forest is most readily observed, and illustrate the two limiting behaviours discussed below. We further assess the robustness of the result to variations in the IGM temperature at mean density, $\mathcal{T}_0$, at $z = 4$. The corresponding temperature-dependent ratios are shown in Fig.~\ref{fig:Result1_Thermals}. \revend

\revstart The separation between the two mixed dark matter treatments grows systematically with redshift. At $k = 0.1\,\mathrm{s\,km^{-1}}$ and $z = 2$, the full Schrödinger--Poisson evolution suppresses the flux power by $10.2$ per cent relative to $\Lambda$CDM, against $7.5$ per cent for the $N$-body approximation with matched initial total-matter density power and random phases, a relative difference of $3.0 \pm 0.6$ per cent. By $z = 4$ the same comparison gives $38.4$ and $21.9$ per cent respectively, a relative difference of $21.2 \pm 0.5$ per cent. Quoted uncertainties are the standard error across transverse groups of skewers within the simulated volume; they do not include variation between independent simulation volumes. The difference decreases towards lower redshift. Its detailed origin depends on the evolving density and velocity fields and their mean-flux-calibrated mapping into transmitted flux; we have not separated these contributions. \revend

\revstart The separation also depends on scale. At $z = 4$ the two treatments differ by $5.3$ per cent at $k = 0.02\,\mathrm{s\,km^{-1}}$, rising to $21.2$ per cent at $k = 0.1$ and $26.3$ per cent at $k = 0.2\,\mathrm{s\,km^{-1}}$, so the divergence develops through the axion Jeans wavenumber, $k_J(z=4)\simeq0.055\,\mathrm{s\,km^{-1}}$ (Eq.~\ref{eq:kjeans}), and keeps growing beyond it. At the largest scales the ratio is additionally affected by the mean-flux calibration. Because the optical depth distribution differs between cosmologies, enforcing a common mean flux requires a different amplitude $A$ in each box. This rescaling alters the shape of the flux probability distribution and not only its mean, and consequently introduces a small offset in the flux power ratio even on scales where the underlying density fields are indistinguishable \citep[cf.][]{Viel2013}. We therefore restrict quantitative claims to intermediate scales. \revend

\revstart We note that the divergence continues to grow beyond $k \simeq 0.1\,\mathrm{s\,km^{-1}}$, but that this is the regime, at $z>4$, in which current data constrain the pressure-smoothing scale and thermal state \citep{Boera2019,Irsic2024}, and in which the peculiar velocity field is sensitive to baryonic response not captured by a dark-matter-only treatment. In hydrodynamical simulations with inhomogeneous reionisation, spatial variations in the thermal broadening and the divergent peculiar-velocity field of overpressurised gas suppress the flux power spectrum by $\sim10$ per cent at $k\sim0.1\,\mathrm{s\,km^{-1}}$ at $4.2\leq z\leq5$ \citep{Molaro2022}. We therefore do not base quantitative claims on this regime, and the differences quoted here should be read as a statement about dark matter kinematics within an idealised mapping, not as a quantitative forecast of the observable difference. \revend

\revstart We focus on $z = 4$ in what follows. This allows us to retain greater sensitivity to small-scale suppression and is closer to the regime used for strong small-scale dark-matter constraints; $z = 2$ lies near the low-redshift limit of the forest, where the absorption signal is weaker. \revend

Fig.~\ref{fig:NLGrowth} demonstrates that the matter power spectra of the two mixed-DM treatments differ by $\sim 25$ per cent at $k \sim 10\,h\,\mathrm{Mpc}^{-1}$ by $z = 4$. The line-of-sight velocity power spectra, shown in Fig.~\ref{fig:SpeedPS}, diverge more strongly. At $z=30$, in the linear regime, the two treatments already differ in velocity power: on small scales the $N$-body box retains $\sim70$ per cent of the $\Lambda$CDM power and the full composite $\sim40$ per cent. This offset is inherited from the initial conditions, because the composite velocity of the full box includes the axion velocity transfer function, which is suppressed beyond the Jeans scale, whereas the $N$-body box is assigned CDM-like velocities (Section~\ref{sec:Method}). Part of the velocity divergence is therefore set at the initial conditions and part is generated by the subsequent evolution. At $z = 4$ the gap has widened, most strongly at the smallest scales: the $N$-body approximation retains $\sim 40$ per cent of the $\Lambda$CDM velocity power on small scales, while the full Schrödinger–Poisson treatment drives the velocity power to near zero beyond the Jeans scale.

\revstart The suppression is sensitive to both the composite velocity construction and differences in the particle sector.  \revend

\revstart The first is the density-weighted construction of the composite velocity field
(Eq. \ref{eq:vlos_composite}). As structure formation proceeds, the CDM
component clusters preferentially into halos and filament cores while the
smoother axion field, supported by quantum pressure, retains a more diffuse
distribution. Where the local axion fraction exceeds its cosmic mean, the composite velocity is weighted towards a component carrying less small-scale kinematic structure. Figure \ref{fig:Result1_Weights} tests the sensitivity to this local weighting: using the global fraction results in reduced \LyA suppression, and using CDM speed information exclusively reduces the signature further. \revend

\revstart The second is that the CDM particles themselves move differently. Repeating
the analysis with the axion velocity contribution taken out entirely, so that the line-of-sight velocity is that of the
$N$-body particles alone, we find the suppression persists: the divergence between the two treatments is reduced but not eliminated. The separate particle-velocity diagnostic in Fig.~\ref{fig:SpeedPS} also shows a kinematic difference in the particle sector, whose evolution is coupled gravitationally to the FDM field. The flux ablation alone does not isolate this contribution, because the density field and initial particle prescription also differ between the two treatments. It establishes that the flux difference is not removed by omitting the axion velocity from the forward model. \revend

\begin{figure}
    \centering
    \includegraphics[width=\linewidth]{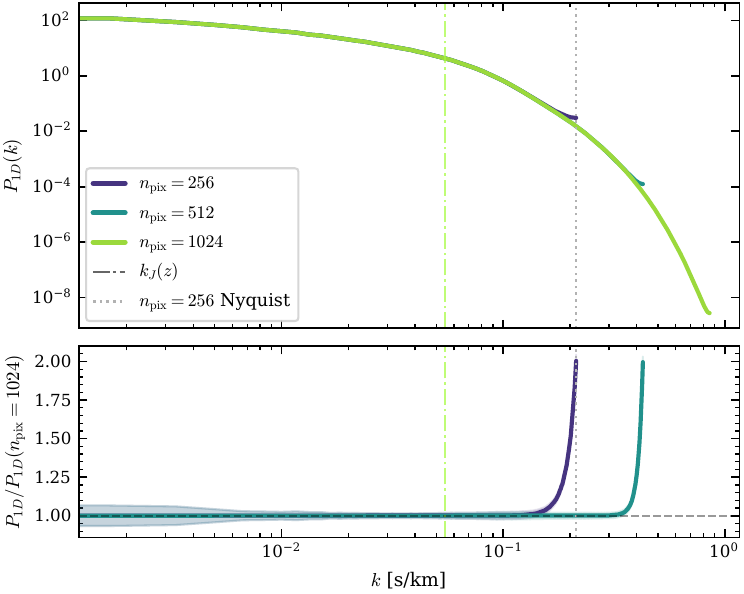}
    \caption{\color{revcolor} \textbf{Top:} The $z=4$ MDM Full \LyA flux power spectrum computed with the optical depth evaluated on output grids of $n_{\rm pix}=256$, $512$, and $1024$ pixels, with the underlying simulation and skewer sampling held fixed. \newline \textbf{Bottom:} Ratio of the $256$- and $512$-pixel spectra to the fiducial $1024$-pixel result, evaluated over their shared $k$-range. The dash-dotted line marks the axion Jeans wavenumber $k_J(z=4)$; the dotted line marks the Nyquist frequency of the coarsest ($n_{\rm pix}=256$) grid. Both ratios remain flat at unity across the shared range and depart only in a narrow band at their own Nyquist limit.}
    \label{fig:NpixAblation}
\end{figure}

\revstart As a check that the reported suppression is not an artefact of the fixed $1024$-pixel output resolution used throughout, we recompute the MDM Full flux power spectrum at $z=4$ with the optical depth evaluated on coarser output grids of $256$ and $512$ pixels, keeping the underlying $1024^3$ simulation and skewer sampling unchanged. Fig.~\ref{fig:NpixAblation} shows the ratio of each coarser-pixel spectrum to the fiducial $1024$-pixel result. The two agree with the fiducial spectrum to better than $1$ per cent up to $k \simeq 0.13\,\mathrm{s\,km^{-1}}$ for the $256$-pixel case ($\sim\!63$ per cent of its own Nyquist frequency) and $k \simeq 0.34\,\mathrm{s\,km^{-1}}$ for the $512$-pixel case, comfortably enclosing the axion Jeans scale at this redshift ($k_J(z=4) \simeq 0.055\,\mathrm{s\,km^{-1}}$, where the two agree with the fiducial to within $0.01$ per cent). Both curves depart sharply from unity only in a narrow band immediately below their respective Nyquist frequencies, outside the range used for the reported intermediate-scale comparisons. We conclude that the suppression signal reported above is not sensitive to the choice of output pixelisation. \revend

\revstart A real-space counterpart of the MDM velocity suppression is shown in
Fig.~\ref{fig:Skewer}, which compares the transmitted flux along a single
sightline in all three simulations. Because the boxes share initial random
phases, the same sightline samples broadly corresponding structures in each, although differing evolution and redshift-space mapping can shift the absorption features. What differs is
their internal structure: the full Schrödinger--Poisson treatment produces
visibly smoother troughs than either the $N$-body approximation or
$\Lambda$CDM, with the sharp, narrow features of the latter partially washed
out. \revend

\section{Conclusions}\label{sec:Conclusion}

\revstart We have investigated the Lyman-$\alpha$ flux power spectrum in mixed FDM--CDM cosmologies using direct Schrödinger--Poisson simulations,  comparing a full wave-mechanical treatment against an $N$-body approximation with matched initial total-matter density power and random phases. Both treatments suppress the flux power spectrum relative to $\Lambda$CDM at scales $k \sim 0.05$--$0.2\,\mathrm{s\,km}^{-1}$, but the full SP treatment produces significantly stronger suppression. Within the adopted dark-matter-only FGPA mapping, the full wave treatment produces a kinematic difference that is not captured by a proxy based only on suppressed initial density power. Determining whether this difference survives neutral-hydrogen weighting, nonlinear baryonic response, and thermal-history marginalisation requires full hydrodynamical simulations. Preliminary tests weighting the velocity diagnostic by the local $n_{\rm HI}$ field suggest the divergence is reduced but still present; a quantitative comparison requires a fully specified weighting and power-spectrum estimator, which we leave to future work. \revend

\begin{figure}
    \centering
    \includegraphics[width=\linewidth]{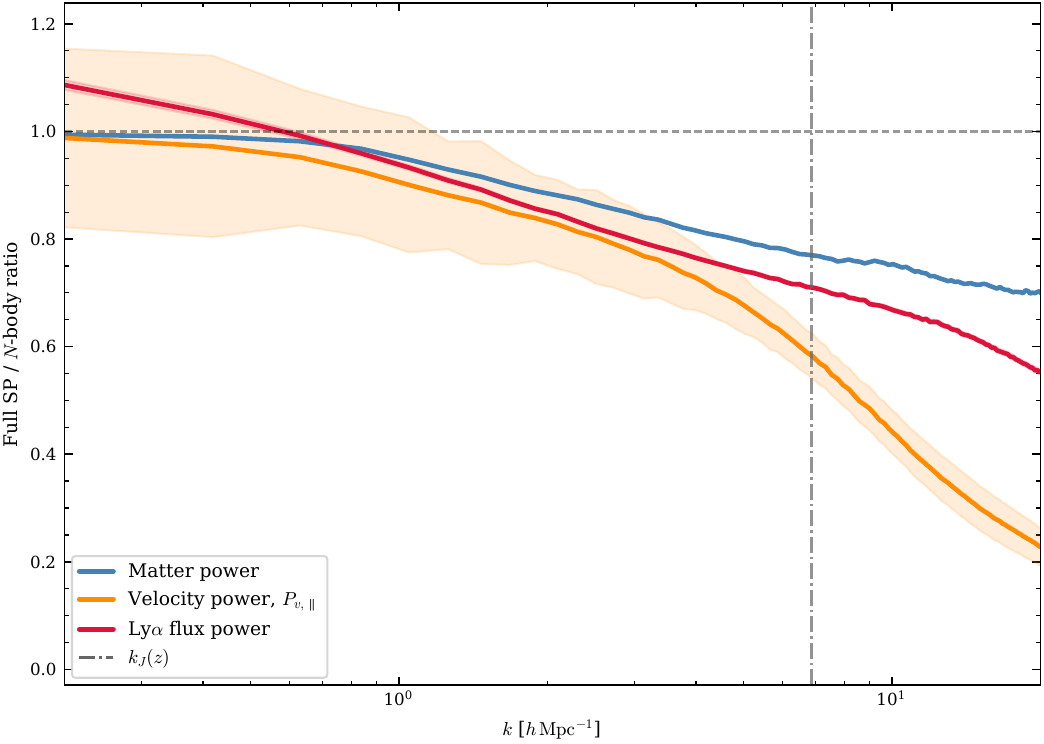}
    \caption{\color{revcolor} Ratio of the full Schrödinger--Poisson MDM calculation to the matched $N$-body approximation at $z=4$, for the matter power spectrum, the line-of-sight velocity power spectrum, and the \LyA\ flux power spectrum. Values below unity indicate additional suppression in the full treatment relative to the particle-only proxy. The velocity power is computed from the unfiltered, unweighted composite field (Section~\ref{sec:NHIWeighting}). Initial total-density power and random phases are matched, but the initial velocity prescriptions differ; the ratios therefore include both the inherited velocity offset and subsequent dynamical differences. The matter power ratio declines only modestly across the range shown ($\sim25$ per cent by $k\sim10\,h\,\mathrm{Mpc}^{-1}$), while the velocity power ratio departs from unity more steeply; the flux power ratio follows an intermediate path between the two. Shaded bands show the propagated standard error on the velocity and flux ratios; this propagation treats the input estimates as independent and does not include covariance from their shared baseline or matched phases. The vertical dash-dotted line marks the axion Jeans wavenumber $k_J(z=4)$.}
    \label{fig:FullVsNbody}
\end{figure}

\revstart The origin of this divergence lies not primarily in the matter power spectrum, which diverges only modestly between the two treatments, but in the
kinematic structure of the velocity field. The full treatment has a suppressed LOS velocity power spectrum, while the $N$-body approximation retains more small-scale velocity power. Fig.~\ref{fig:FullVsNbody} shows this directly: the full-SP-to-$N$-body ratio declines only modestly for the matter power spectrum across the range shown ($\sim25$ per cent by $k\sim10\,h\,\mathrm{Mpc}^{-1}$), falls more steeply for the velocity power spectrum, and the \LyA flux power spectrum tracks an intermediate path between the two, reflecting the partial imprint of the velocity divergence on the forest through the FGPA mapping. This demonstrates sensitivity to the velocity prescription beyond that captured by initial density suppression in the tested proxy. Whether this leads to a bias in observational constraints from particle-based models requires validation with hydrodynamics and thermal-history marginalisation. Our results provide a complementary nonlinear perspective to recent perturbative extensions of mixed dark matter \citep{Verd2025}. \revend

\revstart \citet{NoriMNRAS} used the effective-quantum-pressure code \textsc{Ax-Gadget} to compare pure-FDM hydrodynamical simulations with and without the quantum-pressure force active. They report additional flux-power suppression below $5$ per cent over the scales and masses studied; the inferred mass bound changes from $20.45$ to $21.08\times10^{-22}$ eV, approximately $3$ per cent. Their lightest simulated mass, $m_{22}=2.5$, is $250$ times heavier than our benchmark. Since the axion Jeans wavenumber scales as $m_{22}^{1/2}$, the lower mass considered here places the characteristic scale at a smaller wavenumber. This scaling does not by itself predict the relative flux-suppression amplitudes, which also depend on the FDM fraction, initial velocities and baryonic treatment. Both comparisons find that omitting the respective FDM dynamics reduces the suppression, but the effective quantum-pressure treatment and the full Schrödinger--Poisson calculation need not capture the same nonlinear physics. \revend

We note that our $N$-body component uses one particle per cell. Shot noise from particle deposition contributes stochastic power to the $N$-body velocity field at high $k$, and its net effect on the SP--$N$-body divergence is not straightforward to determine: it may partially mask genuine velocity suppression or artificially inflate small-scale velocity power. Higher particle counts would reduce this ambiguity and are planned for future work.

\revstart Our analysis identifies a kinematic difference between the full wave treatment and the specified particle-only proxy within an idealised FGPA mapping. Extending to a broader parameter space, incorporating full hydrodynamical evolution, and separating initial-velocity effects from subsequent dynamics are needed to quantify its impact on observational constraints. The results motivate testing how the velocity prescription affects Lyman-$\alpha$ predictions in mixed dark matter cosmologies. \revend

\section*{Data Availability}
The simulation data and configuration files underlying this article will be shared upon reasonable request. The version of \texttt{AxioNyx} (1.1.7) used in this work is archived as \citet{AxioNyx_Zenodo}. The bespoke FGPA pipeline, \texttt{AOTS}, is archived as \citet{AOTS_Zenodo}. The modified \texttt{axiECAMB} used for the Newtonian-gauge velocity transfers described in Section~\ref{sec:Method} is archived as \citet{AxiECAMB_Zenodo}. \texttt{MUSIC} is publicly available.

\section*{Acknowledgments}
YW thanks the anonymous referee for the feedback on the analysis pipeline, Jens Niemeyer for discussions, Keir Rogers, Alex Laguë and Dan Grin for their assistance on creating the prototype simulations, and Paolo Codecasa, whose BSc thesis \citep{Codecasa2025} developed an independent FGPA pipeline for related Lyman-$\alpha$ forecasts.
The author also thanks
David Marsh,
Renée Hložek,
Mihir Kulkarni,
Andrew Eberhardt, and
Agata Wisłocka,
for helpful discussions.

The author gratefully acknowledges the computing time granted by the Resource Allocation Board and provided on the supercomputer Emmy at NHR-Nord@Göttingen as part of the NHR infrastructure under grant number nip00084. The visualisations of the cosmological simulations are produced with \texttt{yt} \citep{yt} and \texttt{matplotlib} \citep{MPL}.


\bibliographystyle{mnras}
\bibliography{bibliography}

@article{Viel2013,
  title = {Warm dark matter as a solution to the small scale crisis: New constraints from high redshift Lyman-$\ensuremath{\alpha}$ forest data},
  author = {Viel, Matteo and Becker, George D. and Bolton, James S. and Haehnelt, Martin G.},
  journal = {Phys. Rev. D},
  volume = {88},
  issue = {4},
  pages = {043502},
  numpages = {19},
  year = {2013},
  month = {Aug},
  publisher = {American Physical Society},
  doi = {10.1103/PhysRevD.88.043502},
  url = {https://link.aps.org/doi/10.1103/PhysRevD.88.043502}
}

@article{Onorbe2017,
doi = {10.3847/1538-4357/aa6031},
url = {https://doi.org/10.3847/1538-4357/aa6031},
year = {2017},
month = {mar},
publisher = {The American Astronomical Society},
volume = {837},
number = {2},
pages = {106},
author = {Oñorbe, Jose and Hennawi, Joseph F. and Lukić, Zarija},
title = {Self-consistent Modeling of Reionization in Cosmological Hydrodynamical Simulations},
journal = {The Astrophysical Journal},
}

@article{Planck18,
    author = "{Planck Collaboration}",
    title = "{Planck 2018 results. VI. Cosmological parameters}",
    eprint = "1807.06209",
    archivePrefix = "arXiv",
    primaryClass = "astro-ph.CO",
    doi = "10.1051/0004-6361/201833910",
    journal = "Astron. Astrophys.",
    volume = "641",
    pages = "A6",
    year = "2020",
    note = "[Erratum: Astron.Astrophys. 652, C4 (2021)]"
}

@misc{AxioNyx_Zenodo,
  author       = {Wang, Yourong Frank and Schwabe, Bodo and Behrens, Christoph and Almgren, Ann and Friesen, Brian and Luki{\'c}, Zarija and Sexton, Jean M. and McCorquodale, Peter and Zhang, Weiqun and Myers, Andrew and Beckner, Vince and Schmidt, Wolfram and {Dmitriy} and Zingale, Michael and Weber, Gunther H.},
  title        = {{Sifyrena/axionyx: Axionyx 1.1.7}},
  month        = sep,
  year         = 2026,
  publisher    = {Zenodo},
  version      = {1.1.7},
  doi          = {10.5281/zenodo.22874200},
  url          = {https://doi.org/10.5281/zenodo.22874200},
  howpublished = {Computer software}
}

@mastersthesis{Codecasa2025,
  author = {Codecasa, Paolo Jacopo},
  title  = {Wavelet scattering forecasts for fuzzy dark matter in the Lyman-$\alpha$ forest},
  school = {Georg-August-Universit\"at G\"ottingen},
  year   = {2025},
  type   = {Bachelor's thesis},
}

@misc{AOTS_Zenodo,
  author       = {Wang, Yourong Frank},
  title        = {{Sifyrena/AOTS: Initial Release (Zenodo Compatible)}},
  month        = sep,
  year         = 2026,
  publisher    = {Zenodo},
  version      = {v1.3z},
  doi          = {10.5281/zenodo.22874466},
  url          = {https://doi.org/10.5281/zenodo.22874466},
  howpublished = {Computer software}
}

@misc{AxiECAMB_Zenodo,
  author       = {Wang, Yourong Frank},
  title        = {{Sifyrena/AxiECAMB-vTs: Newtonian-gauge velocity transfers}},
  month        = sep,
  year         = 2026,
  publisher    = {Zenodo},
  doi          = {10.5281/zenodo.22875076},
  url          = {https://doi.org/10.5281/zenodo.22875076},
  howpublished = {Computer software}
}

@preprint{Liu2026MDM,
    author = "Liu, Jianxiang and Gong, Yan and Zhou, Xingchen",
    title = {Lyman-$\alpha$ forest constraints on pure and mixed fuzzy dark matter},
    eprint = "2606.06969",
    archivePrefix = "arXiv",
    primaryClass = "astro-ph.CO",
    month = "6",
    year = "2026"
}

@article{AxiECAMB,
  title = {Accurate method for ultralight axion CMB and matter power spectra},
  author = {Liu, Rayne and Hu, Wayne and Grin, Daniel},
  journal = {Phys. Rev. D},
  volume = {112},
  issue = {2},
  pages = {023513},
  numpages = {28},
  year = {2025},
  month = {Jul},
  publisher = {American Physical Society},
  doi = {10.1103/1z4c-1w7f},
  url = {https://link.aps.org/doi/10.1103/1z4c-1w7f}
}

@article{Lukic2014,
    author = {Lukić, Zarija and Stark, Casey W. and Nugent, Peter and White, Martin and Meiksin, Avery A. and Almgren, Ann},
    title = {The Lyman $\alpha$ forest in optically thin hydrodynamical simulations},
    journal = {Monthly Notices of the Royal Astronomical Society},
    volume = {446},
    number = {4},
    pages = {3697-3724},
    year = {2014},
    month = {12},
    issn = {0035-8711},
    doi = {10.1093/mnras/stu2377},
}

@article{NiemeyerR2020,
title = {Small-scale structure of fuzzy and axion-like dark matter},
journal = {Progress in Particle and Nuclear Physics},
volume = {113},
pages = {103787},
year = {2020},
issn = {0146-6410},
doi = {https://doi.org/10.1016/j.ppnp.2020.103787},
url = {https://www.sciencedirect.com/science/article/pii/S014664102030034X},
author = {Jens C. Niemeyer},
}

@article{Kobayashi2017,
  title = {Lyman-$\ensuremath{\alpha}$ constraints on ultralight scalar dark matter: Implications for the early and late universe},
  author = {Kobayashi, Takeshi and Murgia, Riccardo and De Simone, Andrea and Ir\ifmmode \check{s}\else \v{s}\fi{}i\ifmmode \check{c}\else \v{c}\fi{}, Vid and Viel, Matteo},
  journal = {Phys. Rev. D},
  volume = {96},
  issue = {12},
  pages = {123514},
  numpages = {16},
  year = {2017},
  month = {Dec},
  publisher = {American Physical Society},
  doi = {10.1103/PhysRevD.96.123514},
  url = {https://link.aps.org/doi/10.1103/PhysRevD.96.123514}
}

@article{NoriMNRAS,
    author = "Nori, Matteo and Murgia, Riccardo and Ir\v{s}i\v{c}, Vid and Baldi, Marco and Viel, Matteo",
    title = "{Lyman $\alpha$ forest and non-linear structure characterization in Fuzzy Dark Matter cosmologies}",
    eprint = "1809.09619",
    archivePrefix = "arXiv",
    primaryClass = "astro-ph.CO",
    doi = "10.1093/mnras/sty2888",
    journal = "Mon. Not. Roy. Astron. Soc.",
    volume = "482",
    number = "3",
    pages = "3227--3243",
    year = "2019"
}

@article{Verd2025,
      title={The Effective Field Theory of Large Scale Structure for Mixed Dark Matter Scenarios},
      author={Verdiani, Francesco and Castorina, Emanuele and Salvioni, Ennio and Sefusatti, Emiliano},
      journal={Journal of Cosmology and Astroparticle Physics},
      volume={2026},
      number={01},
      pages={047},
      year={2026},
      doi={10.1088/1475-7516/2026/01/047},
      eprint={2507.08792},
      archivePrefix={arXiv},
      primaryClass={astro-ph.CO},
}

@article{gallagher_coles_2022, 
  title={Evolution of Cosmic Voids in the Schr{\"o}dinger-Poisson Formalism}, 
  volume={5}, 
  DOI={10.21105/astro.2208.13851}, 
  url = {https://arxiv.org/abs/2208.13851},
  journal={The Open Journal of Astrophysics}, 
  author={Gallagher, Aoibhinn and Coles, Peter}, 
  year={2022}, 
  month={Nov}
}

@preprint{Eberhardt2025,
    author = "Eberhardt, Andrew and Ferreira, Elisa G. M.",
    title = "{Ultralight fuzzy dark matter review}",
    eprint = "2507.00705",
    archivePrefix = "arXiv",
    primaryClass = "astro-ph.CO",
    month = "7",
    year = "2025"
}

@article{May2022,
    author = "May, Simon and Springel, Volker",
    title = "{The halo mass function and filaments in full cosmological simulations with fuzzy dark matter}",
    eprint = "2209.14886",
    archivePrefix = "arXiv",
    primaryClass = "astro-ph.CO",
    doi = "10.1093/mnras/stad2031",
    journal = "Mon. Not. Roy. Astron. Soc.",
    volume = "524",
    number = "3",
    pages = "4256--4274",
    year = "2023"
}

@ARTICLE{MUSIC,
       author = {{Hahn}, Oliver and {Abel}, Tom},
        title = "{Multi-scale initial conditions for cosmological simulations}",
      journal = {\mnras},
         year = 2011,
        month = aug,
       volume = {415},
       number = {3},
        pages = {2101-2121},
          doi = {10.1111/j.1365-2966.2011.18820.x},
archivePrefix = {arXiv},
       eprint = {1103.6031},
 primaryClass = {astro-ph.CO},
       adsurl = {https://ui.adsabs.harvard.edu/abs/2011MNRAS.415.2101H}
}

@article{Gnedin1998,
    author = {Gnedin, Nickolay Y. and Hui, Lam},
    title = {Probing the Universe with the Ly$\alpha$ forest --- I. Hydrodynamics of the low-density intergalactic medium},
    journal = {Monthly Notices of the Royal Astronomical Society},
    volume = {296},
    number = {1},
    pages = {44-55},
    year = {1998},
    month = {05},
    issn = {0035-8711},
    doi = {10.1046/j.1365-8711.1998.01249.x},
    url = {https://doi.org/10.1046/j.1365-8711.1998.01249.x},
}

@article{Lague2023,
    author = {Lagu{\"e}, Alex and Schwabe, Bodo and Hlo{\v{z}}ek, Ren{\'e}e and Marsh, David J. E. and Rogers, Keir K.},
    title = "{Cosmological simulations of mixed ultralight dark matter}",
    eprint = "2310.20000",
    archivePrefix = "arXiv",
    primaryClass = "astro-ph.CO",
    reportNumber = "KCL-PH-TH/2023-57",
    doi = "10.1103/PhysRevD.109.043507",
    journal = "Phys. Rev. D",
    volume = "109",
    number = "4",
    pages = "043507",
    year = "2024"
}

@article{Hernquist96,
	adsurl = {https://ui.adsabs.harvard.edu/abs/1996ApJ...457L..51H},
	archiveprefix = {arXiv},
	author = {{Hernquist}, Lars and {Katz}, Neal and {Weinberg}, David H. and {Miralda-Escud{\'e}}, Jordi},
	doi = {10.1086/309899},
	eprint = {astro-ph/9509105},
	journal = {ApJ Letters},
	month = feb,
	pages = {L51},
	primaryclass = {astro-ph},
	title = {{The Lyman-Alpha Forest in the Cold Dark Matter Model}},
	volume = {457},
	year = 1996}

@misc{AxionCAMB22,
	adsurl = {https://ui.adsabs.harvard.edu/abs/2022ascl.soft03026G},
	archiveprefix = {ascl},
	author = {{Grin}, Daniel and {Marsh}, David J.~E. and {Hlozek}, Renee},
	eid = {ascl:2203.026},
	eprint = {2203.026},
	howpublished = {Astrophysics Source Code Library, record ascl:2203.026},
	month = mar,
	pages = {ascl:2203.026},
	title = {{axionCAMB: Modification of the CAMB Boltzmann code}},
	year = 2022}

@article{Multifield,
	author = {Gosenca, Mateja and Eberhardt, Andrew and Wang, Yourong and Eggemeier, Benedikt and Kendall, Emily and Zagorac, J. Luna and Easther, Richard},
	doi = {10.1103/PhysRevD.107.083014},
	issue = {8},
	journal = {Phys. Rev. D},
	month = {Apr},
	numpages = {11},
	pages = {083014},
	publisher = {American Physical Society},
	title = {Multifield ultralight dark matter},
	url = {https://link.aps.org/doi/10.1103/PhysRevD.107.083014},
	volume = {107},
	year = {2023}}

@article{Irsic2017,
	author = {Ir{\v{s}}i{\v{c}}, Vid and Viel, Matteo and Haehnelt, Martin G. and Bolton, James S. and Becker, George D.},
	doi = {10.1103/PhysRevLett.119.031302},
	issn = {0031-9007},
	journal = {Phys. Rev. Lett.},
	month = {jul},
	number = {3},
	pages = {031302},
	pmid = {28777592},
	title = {{First Constraints on Fuzzy Dark Matter from Lyman-$\alpha$ Forest Data and Hydrodynamical Simulations}},
	url = {http://link.aps.org/doi/10.1103/PhysRevLett.119.031302},
	volume = {119},
	year = {2017}}

@article{Hui2017,
	archiveprefix = {arXiv},
	arxivid = {1610.08297},
	author = {Hui, Lam and Ostriker, Jeremiah P. and Tremaine, Scott and Witten, Edward},
	doi = {10.1103/PhysRevD.95.043541},
	eprint = {1610.08297},
	issn = {24700029},
	journal = {Phys. Rev. D},
	number = {4},
	title = {{Ultralight scalars as cosmological dark matter}},
	volume = {95},
	year = {2017}}

@article{Schwabe2020,
 title = {Simulating mixed fuzzy and cold dark matter},
  author = {Schwabe, Bodo and Gosenca, Mateja and Behrens, Christoph and Niemeyer, Jens C. and Easther, Richard},
  journal = {Phys. Rev. D},
  volume = {102},
  issue = {8},
  pages = {083518},
  numpages = {10},
  year = {2020},
  month = {Oct},
  publisher = {American Physical Society},
  doi = {10.1103/PhysRevD.102.083518},
  url = {https://link.aps.org/doi/10.1103/PhysRevD.102.083518}
}

@ARTICLE{DESILyADR1,
       author = {{Chaves-Montero}, J. and {Font-Ribera}, A. and {McDonald}, P. and {Armengaud}, E. and {Chebat}, D. and {Garcia-Quintero}, C. and {Kara{\c{c}}ayl{\i}}, N.~G. and {Ravoux}, C. and {Satyavolu}, S. and {Sch{\"o}neberg}, N. and {Walther}, M. and {Aguilar}, J. and {Ahlen}, S. and {Bailey}, S. and {Bianchi}, D. and {Brooks}, D. and {Claybaugh}, T. and {Cuceu}, A. and {de la Macorra}, A. and {Doel}, P. and {Ferraro}, S. and {Forero-Romero}, J.~E. and {Gazta{\~n}aga}, E. and {Gontcho}, S. Gontcho A and {Gonzalez-Morales}, A.~X. and {Gutierrez}, G. and {Guy}, J. and {Hahn}, C. and {Herrera-Alcantar}, H.~K. and {Honscheid}, K. and {Ishak}, M. and {Joyce}, R. and {Juneau}, S. and {Kirkby}, D. and {Kremin}, A. and {Lahav}, O. and {Lamman}, C. and {Landriau}, M. and {Le Goff}, J.~M. and {Le Guillou}, L. and {Leauthaud}, A. and {Levi}, M.~E. and {Manera}, M. and {Martini}, P. and {Meisner}, A. and {Miquel}, R. and {Moustakas}, J. and {Nadathur}, S. and {Niz}, G. and {Palanque-Delabrouille}, N. and {Percival}, W.~J. and {Prada}, F. and {P{\'e}rez-R{\`a}fols}, I. and {Rossi}, G. and {Sanchez}, E. and {Schlegel}, D. and {Schubnell}, M. and {Seo}, H. and {Silber}, J. and {Sprayberry}, D. and {Tan}, T. and {Tarl{\'e}}, G. and {Weaver}, B.~A. and {Y{\`e}che}, C. and {Zhou}, R. and {Zou}, H.},
        title = "{Cosmological analysis of the DESI DR1 Lyman alpha 1D power spectrum}",
      journal = {arXiv e-prints},
         year = 2026,
        month = jan,
          eid = {arXiv:2601.21432},
        pages = {arXiv:2601.21432},
          doi = {10.48550/arXiv.2601.21432},
archivePrefix = {arXiv},
       eprint = {2601.21432},
 primaryClass = {astro-ph.CO},
       adsurl = {https://ui.adsabs.harvard.edu/abs/2026arXiv260121432C}
}

@article{Veltmaat2018,
	author = {J. Veltmaat and { Jens} { Niemeyer} and B. Schwabe},
	doi = {10.1103/physrevd.98.043509},
	journal = {Phys. Rev. D},
	month = aug,
	number = {4},
	publisher = {American Physical Society ({APS})},
	title = {{Formation and structure of ultralight bosonic dark matter halos}},
	volume = {98},
	year = {2018}}

@article{RogersPeiris2021,
    author = "Rogers, Keir K. and Peiris, Hiranya V.",
    title = "{Strong Bound on Canonical Ultralight Axion Dark Matter from the Lyman-Alpha Forest}",
    eprint = "2007.12705",
    archivePrefix = "arXiv",
    primaryClass = "astro-ph.CO",
    doi = "10.1103/PhysRevLett.126.071302",
    journal = "Phys. Rev. Lett.",
    volume = "126",
    number = "7",
    pages = "071302",
    year = "2021"
}

@article{Almgren2013,
	author = {A. S. Almgren and . B. Bell and M. J. Lijewski and Z. Luki{\'{c}} and E. Van Andel},
	doi = {10.1088/0004-637x/765/1/39},
	journal = {ApJ},
	month = feb,
	number = {1},
	pages = {39},
	publisher = {{IOP} Publishing},
	title = {{Nyx: A Massively Parallel AMR Code for Computational Cosmology}},
	volume = {765},
	year = {2013}}

@software{Almgren2019,
	author = {Almgren, Ann and Beckner, Vince and Blaschke, Johannes and Chan, Cy and Day, Marcus and Friesen, Brian and Gott, Kevin and Graves, Daniel and Katz, Maximilian and Myers, Andrew and Nguyen, Tan and Nonaka, Andrew and Rosso, Michele and Williams, Sam and Zhang, Weiqun and Zingale, Michael},
	doi = {10.5281/zenodo.2555438},
	publisher = {Zenodo},
	title = {AMReX-Codes/amrex: AMReX 19.09},
	year = {2019}}

@article{Hu2000,
	author = {Hu, Wayne and Barkana, Rennan and Gruzinov, Andrei},
	doi = {10.1103/physrevlett.85.1158},
	issn = {1079-7114},
	journal = {Phys. Rev. Letters},
	month = {Aug},
	number = {6},
	pages = {1158-1161},
	publisher = {American Physical Society (APS)},
	title = {Fuzzy Cold Dark Matter: The Wave Properties of Ultralight Particles},
	url = {http://dx.doi.org/10.1103/PhysRevLett.85.1158},
	volume = {85},
	year = {2000}}

@Article{MPL,
  Author    = {Hunter, J. D.},
  Title     = {Matplotlib: A 2D graphics environment},
  Journal   = {Computing in Science \& Engineering},
  Volume    = {9},
  Number    = {3},
  Pages     = {90--95},
  publisher = {IEEE COMPUTER SOC},
  doi       = {10.1109/MCSE.2007.55},
  year      = 2007
}

@ARTICLE{yt,
   author = {{Turk}, M.~J. and {Smith}, B.~D. and {Oishi}, J.~S. and {Skory}, S. and
{Skillman}, S.~W. and {Abel}, T. and {Norman}, M.~L.},
    title = "{yt: A Multi-code Analysis Toolkit for Astrophysical Simulation Data}",
  journal = {The Astrophysical Journal Supplement Series},
archivePrefix = "arXiv",
   eprint = {1011.3514},
 primaryClass = "astro-ph.IM",
     year = 2011,
    month = jan,
   volume = 192,
      eid = {9},
    pages = {9},
      doi = {10.1088/0067-0049/192/1/9},
   adsurl = {http://adsabs.harvard.edu/abs/2011ApJS..192....9T}
}

@article{Nori2018,
	adsurl = {https://ui.adsabs.harvard.edu/abs/2018MNRAS.478.3935N},
	archiveprefix = {arXiv},
	author = {{Nori}, Matteo and {Baldi}, Marco},
	doi = {10.1093/mnras/sty1224},
	eprint = {1801.08144},
	journal = {Mon. Not. R. Astron. Soc.},
	month = {Aug},
	number = {3},
	pages = {3935-3951},
	primaryclass = {astro-ph.CO},
	title = {{AX-GADGET: a new code for cosmological simulations of Fuzzy Dark Matter and Axion models}},
	volume = {478},
	year = {2018}}

@article{Li2019,
	adsurl = {https://ui.adsabs.harvard.edu/abs/2019PhRvD..99f3509L},
	archiveprefix = {arXiv},
	author = {{Li}, Xinyu and {Hui}, Lam and {Bryan}, Greg L.},
	doi = {10.1103/PhysRevD.99.063509},
	eid = {063509},
	eprint = {1810.01915},
	journal = {Phys. Rev. D},
	month = {Mar},
	number = {6},
	pages = {063509},
	primaryclass = {astro-ph.CO},
	title = {{Numerical and perturbative computations of the fuzzy dark matter model}},
	volume = {99},
	year = {2019}}

@article{Armengaud2017,
	adsurl = {https://ui.adsabs.harvard.edu/abs/2017MNRAS.471.4606A},
	archiveprefix = {arXiv},
	author = {{Armengaud}, Eric and {Palanque-Delabrouille}, Nathalie and {Y{\`e}che}, Christophe and {Marsh}, David J.~E. and {Baur}, Julien},
	doi = {10.1093/mnras/stx1870},
	eprint = {1703.09126},
	journal = {Mon. Not. R. Astron. Soc.},
	month = {Nov},
	number = {4},
	pages = {4606-4614},
	primaryclass = {astro-ph.CO},
	title = {{Constraining the mass of light bosonic dark matter using SDSS Lyman-{\ensuremath{\alpha}} forest}},
	volume = {471},
	year = {2017}}

@article{Kulkarni2022,
	adsurl = {https://ui.adsabs.harvard.edu/abs/2022ApJ...941L..18K},
	archiveprefix = {arXiv},
	author = {{Kulkarni}, {Mihir} and {Visbal}, Eli and {Bryan}, Greg L. and {Li}, Xinyu},
	doi = {10.3847/2041-8213/aca47c},
	eid = {L18},
	eprint = {2210.11515},
	journal = {ApJ Letters},
	month = dec,
	number = {1},
	pages = {L18},
	primaryclass = {astro-ph.GA},
	title = {{If Dark Matter is Fuzzy, the First Stars Form in Massive Pancakes}},
	volume = {941},
	year = 2022}

@article{Boera2019,
  author  = {Boera, Elisa and Becker, George D. and Bolton, James S. and Nasir, Fahad},
  title   = {Revealing reionization with the thermal history of the intergalactic medium: new constraints from the {Lyman-$\alpha$} flux power spectrum},
  journal = {The Astrophysical Journal},
  volume  = {872},
  pages   = {101},
  year    = {2019},
  doi     = {10.3847/1538-4357/aafee4},
  eprint  = {1809.06980},
  archivePrefix = {arXiv},
  primaryClass  = {astro-ph.CO}
}

@article{Irsic2024,
  author  = {Ir{\v{s}}i{\'c}, Vid and Viel, Matteo and Haehnelt, Martin G. and Bolton, James S. and Molaro, Margherita and Puchwein, Ewald and Boera, Elisa and Becker, George D. and Gaikwad, Prakash and Keating, Laura C. and Kulkarni, Girish},
  title   = {Unveiling dark matter free-streaming at the smallest scales with high redshift {Lyman-$\alpha$} forest},
  journal = {Physical Review D},
  volume  = {109},
  pages   = {043511},
  year    = {2024},
  doi     = {10.1103/PhysRevD.109.043511},
  eprint  = {2309.04533},
  archivePrefix = {arXiv},
  primaryClass  = {astro-ph.CO}
}

@article{Molaro2022,
  author  = {Molaro, Margherita and Ir{\v{s}}i{\'c}, Vid and Bolton, James S. and Keating, Laura C. and Puchwein, Ewald and Gaikwad, Prakash and Haehnelt, Martin G. and Kulkarni, Girish and Viel, Matteo},
  title   = {The effect of inhomogeneous reionisation on the {Lyman-$\alpha$} forest power spectrum at redshift $z>4$: implications for thermal parameter recovery},
  journal = {Monthly Notices of the Royal Astronomical Society},
  volume  = {509},
  pages   = {6119},
  year    = {2022},
  doi     = {10.1093/mnras/stab3416},
  eprint  = {2109.06897},
  archivePrefix = {arXiv},
  primaryClass  = {astro-ph.CO}
}


\bsp	
\label{lastpage}
\end{document}